\documentclass[12pt]{article}
\usepackage{booktabs}

\usepackage{threeparttable}
\usepackage[colorlinks,bookmarksopen,bookmarksnumbered,citecolor=blue,urlcolor=blue,linkcolor=blue]{hyperref}
\usepackage{float}
\usepackage{verbatim} 
\usepackage{amsmath}
\usepackage{enumerate}

\usepackage{graphicx}
\graphicspath{{Pictures/}}
\allowdisplaybreaks[4]

\def\spacingset#1{\renewcommand{\baselinestretch}%
	{#1}\small\normalsize} \spacingset{1}

\usepackage{amsthm,amsmath,enumerate,amsbsy,amsfonts,amssymb,mathabx,amscd,graphicx,algorithm, indentfirst}
\usepackage{natbib}
\usepackage{geometry}
\usepackage{authblk, caption, subcaption}
\usepackage{mathrsfs}
\usepackage{epsfig}

\newtheorem{theorem}{Theorem}

\newtheorem{condition}{Condition}
\newtheorem{assumption}{Assumption}[section]

\usepackage{multirow}

\makeatletter
\newcommand*{\rom}[1]{\expandafter\@slowromancap\romannumeral #1@}
\makeatother

\makeatletter

\makeatletter
\DeclareRobustCommand\widecheck[1]{{\mathpalette\@widecheck{#1}}}
\def\@widecheck#1#2{%
	\setbox\z@\hbox{\m@th$#1#2$}%
	\setbox\tw@\hbox{\m@th$#1%
		\widehat{%
			\vrule\@width\z@\@height\ht\z@
			\vrule\@height\z@\@width\wd\z@}$}%
	\dp\tw@-\ht\z@
	\@tempdima\ht\z@ \advance\@tempdima2\ht\tw@ \divide\@tempdima\thr@@
	\setbox\tw@\hbox{%
		\raise\@tempdima\hbox{\scalebox{1}[-1]{\lower\@tempdima\box
				\tw@}}}%
	{\ooalign{\box\tw@ \cr \box\z@}}}
\makeatother
\RequirePackage[colorlinks,citecolor=blue,urlcolor=blue]{hyperref}

\renewcommand{\baselinestretch}{1.6}
\providecommand{\keywords}[1]{\textbf{\textit{Keywords: }} #1}
\newcommand{\blind}{1}

\begin{document}
	\if1\blind
	{
		\title{\bf \Large
	Matrix GARCH Model: Inference and Application
		}
		\author[1]{Cheng Yu}
		\author[1]{Dong Li}
		\author[2]{Feiyu Jiang}
            \author[3]{Ke Zhu}
	    \affil[1]{\it Center for Statistical Science, Tsinghua University, Beijing 100084, China}
       \affil[2]{\it Department of Statistics and Data Science, School of Management, Fudan University, Shanghai, China}
       \affil[3]{\it Department of Statistics and Actuarial Science, University of Hong Kong, Hong Kong}
		\setcounter{Maxaffil}{0}
		
		\renewcommand\Affilfont{\itshape\small}
		\date{\vspace{-5ex}}
		\maketitle
	} \fi
	\if0\blind
	{
		\bigskip
		\bigskip
		\bigskip
		\begin{center}
			{\Large\bf }
		\end{center}
		\medskip
	} \fi

\setcounter{Maxaffil}{0}
\renewcommand\Affilfont{\itshape\small}
\begin{abstract}
Matrix-variate time series data are largely available in applications.
However, no attempt has been made to study their conditional heteroskedasticity that
is often observed in economic and financial data. To address this gap,
we propose a novel matrix generalized autoregressive conditional heteroskedasticity (GARCH) model to
capture the dynamics of conditional row and column covariance matrices of matrix time series.
The key innovation of the matrix GARCH model is the use of a univariate GARCH specification for the trace of conditional row
or column covariance matrix, which allows for the identification of conditional row and column covariance matrices. Moreover, we introduce a quasi maximum likelihood estimator (QMLE) for model estimation
and develop a portmanteau test for model diagnostic checking. Simulation studies are conducted to
assess the finite-sample performance of the QMLE and portmanteau test.
To handle large dimensional matrix time series, we also propose a matrix factor GARCH model.
Finally, we demonstrate the superiority of the matrix GARCH and matrix factor GARCH models over existing multivariate GARCH-type models in volatility forecasting and portfolio allocations using three applications on credit default swap prices, global stock sector indices, and future prices.
\end{abstract}


\newpage

\section{Introduction}
\label{sec:intro}

The increasing complexity of time series data has led to the collection of data that extend beyond the conventional vectorial mode. When time series data are observed at the intersections of two classifications, they naturally form a matrix structure. Examples of such data include economic indicator data \citep{chen2021autoregressive}, international trade flow data \citep{chen2022tradenetwork}, global stock market data \citep{chen2023testing}, and many others. To study the conditional mean of matrix time series, \cite{chen2021autoregressive} propose a matrix autoregressive (MAR) model. The estimation of MAR model is tractable when the dimension of matrix time series is small, however, its implementation tends to be computationally infeasible when the dimension of matrix time series is large. To handle large dimensional matrix time series, \cite{Xiao2022reduced} propose a reduced rank MAR model to achieve dimension reduction.
Their model is distinct from matrix factor models, which aim to learn a much smaller dimensional
 latent matrix factor that is feasible for modelling or prediction. For a burgeoning literature on matrix factor models,
 one can refer to \cite{Wang2019Factor}, \cite{chen2019constrained}, \cite{chen2021statistical}, \cite{YU2022projected}, \cite{zhang2022modeling}, and \cite{chang2021modelling}.

\begin{figure}[htbp]
	\centering
	\includegraphics[height=6.5cm, width=17cm]{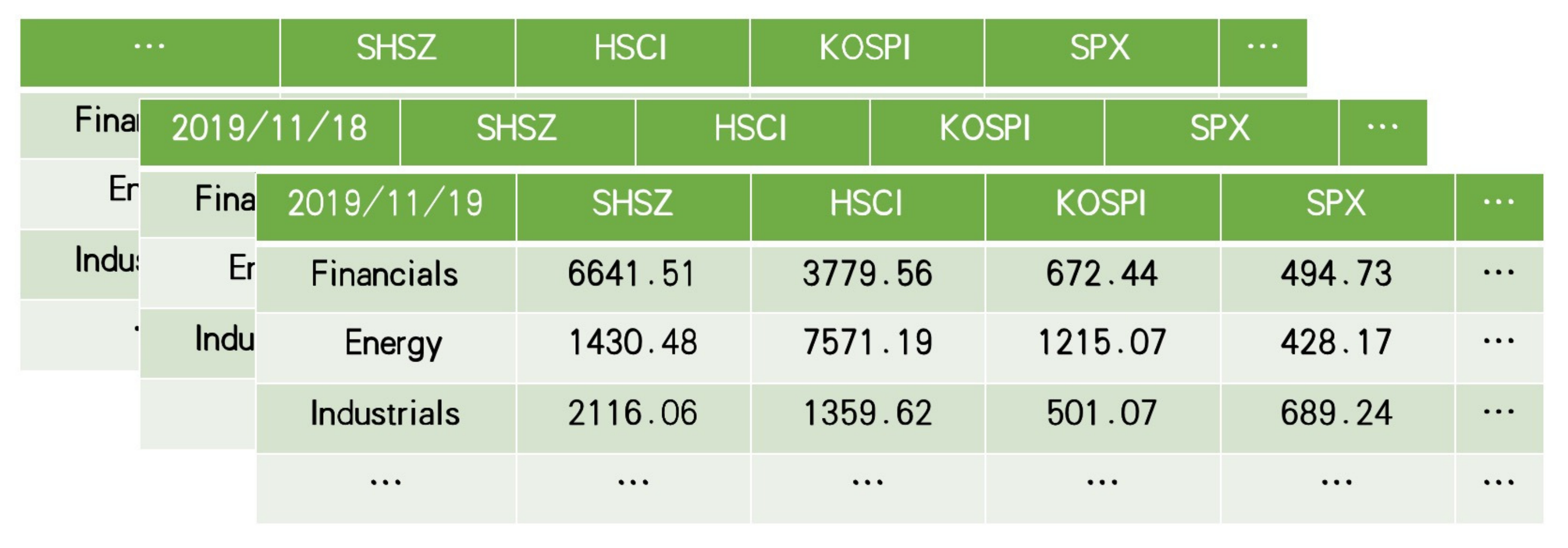}
	\caption{A real example of matrix-variate observations consisting of sector indices in different international stock markets. Data source: Yahoo Finance and Bloomberg.}
	\label{Matrix_Index}
\end{figure}

To date, the aforementioned works focus solely on modelling conditional mean of matrix time series, without accounting for a prevalent
fact that many economic and financial time series data exhibit conditional heteroskedasticity (\citealp{engle1982autoregressive}). As an illustrating example, Figure \ref{Matrix_Index} plots a time series list of tables recording the sector indices in various international stock markets. The returns of these sector indices naturally form an $m\times n$ matrix time series, where $m$ and $n$ are the numbers of sectors (rows) and markets (columns), respectively. For this return matrix time series, we find that the sample autocorrelations of each entry series are insignificant, suggesting the absence of dynamic structure in  its conditional mean. However, we observe that the sample autocorrelations of each squared entry series are very significant. This is a clear indication of the conditional heteroskedasticity, a characteristic cannot be captured by matrix time series models that only account for the conditional mean.

Since the seminal works of \cite{engle1982autoregressive} and \cite{bollerslev1986generalized}, the generalized autoregressive conditional heteroskedasticity (GARCH) model and its variants have become benchmark/standard tools for studying the conditional heteroskedasticity of time series. Along this line, one can transform an $m\times n$ matrix time series into an $mn$-dimensional vector time series by the vectorization operator, and then investigate the conditional heteroskedasticity of this vector time series via multivariate GARCH-type models, such as the
 constant conditional correlation (CCC) model (\citealp{bollerslev1990modelling}),
 the BEKK model (\citealp{engle1995multivariate}), the dynamic conditional correlation (DCC) model (\citealp{engle2002dynamic}), and many others.
See \cite{bauwens2006multivariate} and \cite{francq2019garch} for comprehensive surveys on multivariate GARCH-type models.
However, due to such a vectorization operation, the multivariate GARCH-type models have two major drawbacks to study the conditional heteroskedasticity of matrix time series. First, they bring a damage to the original matrix data structure, making it difficult to  informatively  interpret the conditional row and column covariance matrices.
Second, they need to estimate model parameters of order $O(m^2n^2)$, so the related estimation becomes computationally challenging even for the moderate values of $m$ and $n$.

In this paper, we first propose a new matrix GARCH model to  tackle the aforementioned limitations.
Our matrix GARCH model applies two different BEKK specifications to capture the conditional
row and column covariance matrices, respectively, and at the same time, it adopts a
univariate GARCH specification to depict the trace of conditional row or column covariance
matrix. Formally, the matrix GARCH model is related to the existing BEKK and univariate
GARCH models, and its trace specification is essential to identifying the conditional row and
column covariance matrices. Compared with the aforementioned multivariate GARCH-type
models, the matrix GARCH model offers the advantage of directly interpreting the conditional row and column covariance matrices, while also reducing computational complexity with fewer parameters of order $O(m^2+n^2)$.

Next, in order to provide statistical inference for the proposed matrix GARCH model, we develop a
 quasi maximum likelihood estimator (QMLE) of its parameters and a portmanteau test for its diagnostic checking. Under certain conditions, the asymptotics of the QMLE and portmanteau test are established. Further, we introduce  a new matrix factor GARCH model to handle the situations when either $m$ or $n$ is large. Finally, we assess the finite-sample performance of the QMLE and portmanteau tests by simulation studies, and provide three applications on credit default swap prices, global stock sector indices, and future prices to
demonstrate the advantage of our matrix GARCH and matrix factor GARCH models over existing multivariate GARCH-type models
in volatility forecasting and portfolio allocation.

%
%

The remaining paper is organized as follows. Section \ref{sec:method} introduces the specification of matrix GARCH model.
 Section \ref{sec.estimation} studies the QMLE with its asymptotics for the model. Section \ref{sec: Model Checking} proposes the portmanteau test for the model diagnostic checking. The matrix factor GARCH model is formed in Section \ref{sec: Matrix Factor GARCH}. Simulation results are reported in Section \ref{sec.sim}. Three applications are presented in Section \ref{sec.real}. Concluding remarks are offered in Section \ref{sec_concluding}.
The proofs of all theorems are deferred to the supplementary materials.

Throughout the paper, $\mathbb{R}$ is a real line, $\mathbf{I}_{p}$ is the $p\times p$ identity matrix,
$\mathbf{1}_p$ is the $p$-dimensional vector of ones; ``$\stackrel{p}{\to}$'' and ``$\stackrel{d}{\to}$'' denote the convergence in probability and in distribution, respectively.
For a square matrix $\mathbf{X}\in \mathbb{R}^{p\times p}$, $\text{tr}(\mathbf{X})$, $|\mathbf{X}|$,
 and $\rho(\mathbf{X})$ are its trace, determinant, and  spectral radius, respectively.
For a rectangular matrix $\mathbf{X}\in \mathbb{R}^{p\times q}$,  $\mathbf{X}'$, $\text{vec}(\mathbf{X})$,
$\|\mathbf{X}\|:=\sqrt{\mathrm{tr}(\mathbf{X}'\mathbf{X})}$, and $\|\mathbf{X}\|_{spec}:=\sqrt{\rho(\mathbf{X}'\mathbf{X})}$ are
its transpose, vectorization, Frobenius norm, and spectral norm, respectively.

\section{Matrix GARCH Model}\label{sec:method}

Consider a time series of length $T$, in which at each time point $t$, an $m \times n$ matrix $\mathbf{X}_t$ is observed.
Let $\mathcal{F}_{t}=\sigma(\mathbf{X}_s; s\leq t)$ be the natural filtration containing all available information up to time $t$.
To accommodate the matrix structure of $\mathbf{X}_t$, we study its dynamics of conditional heteroskedasticity via the following matrix time series model
\begin{equation}\label{Eq2.2}
\mathbf{X}_t = \mathbf{U}_t^{1/2} \mathbf{Z}_t \mathbf{V}_t^{1/2},\,\,\,t=1,...,T,
\end{equation}
where $\mathbf{U}_t\in\mathcal{F}_{t-1}$ and $\mathbf{V}_t\in\mathcal{F}_{t-1}$ are
$m\times m$ and $n\times n$ positive definite random matrices, respectively, and $\{\mathbf{Z}_t\}_{t=1}^{T}$ is a sequence of independent and identically distributed (i.i.d.) $m\times n$ random matrix innovations satisfying  $E(\mathbf{Z}_t)=\mathbf{0}$ and
 $E\left[\text{vec}(\mathbf{Z}_t) \text{vec}(\mathbf{Z}_t)'\right]=\mathbf{I}_{mn}$.

 Clearly, model \eqref{Eq2.2} is motivated by the conventional multivariate GARCH models. The two elements, namely,   $\mathbf{U}_t$ and $\mathbf{V}_t$, are key to
studying the conditional row and column covariance matrices of $\mathbf{X}_t$.
In fact,  the conditional row covariance matrix of $\mathbf{X}_t$ in (\ref{Eq2.2}) is
  \begin{align}\label{row_cov}
 E\left(\mathbf{X}_t\mathbf{X}_t'|\mathcal{F}_{t-1}\right)=\text{tr}(\mathbf{V}_t)\mathbf{U}_t,
 \end{align}
by noting that
$E\left(\mathbf{Z}_t\mathbf{V}_t\mathbf{Z}_t'|\mathcal{F}_{t-1}\right)=\text{tr}(\mathbf{V}_t)\mathbf{I}_m$. Similarly,
the conditional column covariance matrix of $\mathbf{X}_t$ in (\ref{Eq2.2}) is
 \begin{align}\label{column_cov}
E\left(\mathbf{X}_t'\mathbf{X}_t|\mathcal{F}_{t-1}\right)=\text{tr}(\mathbf{U}_t)\mathbf{V}_t.
 \end{align}
 The results of \eqref{row_cov} and \eqref{column_cov} demonstrate that the conditional row and column covariances of $\mathbf{X}_t$ are entangled through  $\mathbf{U}_t$ and $\mathbf{V}_t$. However, since $\text{tr}(\mathbf{V}_t)\mathbf{U}_t=\text{tr}(a\mathbf{V}_t)[\mathbf{U}_t/a]$ for any $a\neq0$, $\mathbf{U}_t$ and $\mathbf{V}_t$ can not be identified without a constraint on
 $\text{tr}(\mathbf{U}_t)$ or $\text{tr}(\mathbf{V}_t)$.


To tackle the identification issue of $\mathbf{U}_t$ and $\mathbf{V}_t$, we model $\mathbf{U}_t$ and $\mathbf{V}_t$ by
\begin{align}
\mathbf{U}_t &= \frac{\mathbf{S}_{1,t}}{\text{tr}(\mathbf{S}_{1,t})} y_t,\label{Matrix_volatility_U}\\
\mathbf{V}_t &= \frac{\mathbf{S}_{2,t}}{\text{tr}(\mathbf{S}_{2,t})},\label{Matrix_volatility_V}
\end{align}
where $\mathbf{S}_{1,t}\in\mathcal{F}_{t-1}$ and $\mathbf{S}_{2,t}\in\mathcal{F}_{t-1}$ are $m\times m$ and $n\times n$ positive definite random matrices, respectively, $y_t\in\mathcal{F}_{t-1}$ is a positive random variable, and they satisfy
\begin{align}
\mathbf{S}_{1,t} &= \mathbf{A}_0 \mathbf{A}_0' + \mathbf{A}_1\mathbf{X}_{t-1} \mathbf{X}_{t-1}' \mathbf{A}_1' + \mathbf{A}_2 \mathbf{S}_{1,t-1} \mathbf{A}_2', \label{MGARCH_1}\\
\mathbf{S}_{2,t} &= \mathbf{B}_0 \mathbf{B}_0' + \mathbf{B}_1\mathbf{X}_{t-1}' \mathbf{X}_{t-1} \mathbf{B}_1' + \mathbf{B}_2 \mathbf{S}_{2,t-1} \mathbf{B}_2', \label{MGARCH_2}\\
y_t &= w + \alpha\, \text{tr}(\mathbf{X}_{t-1} \mathbf{X}_{t-1}') + \beta y_{t-1}.\label{MGARCH_3}
\end{align}
Here,  $\mathbf{A}_0$, $\mathbf{A}_1$, and $\mathbf{A}_2$ are $m\times m$ parameter matrices,  $\mathbf{B}_0$, $\mathbf{B}_1$, and $\mathbf{B}_2$ are $n\times n$ parameter matrices, $\mathbf{A}_0$ and $\mathbf{B}_0$ are lower triangular matrices with non-negative diagonal elements and
\begin{align}\label{A_0_B_0}
 \mathbf{A}_{0,11} = \mathbf{B}_{0,11} = 1,
\end{align}
and the scale parameters $w > 0$, $\alpha \ge 0$, and $\beta \ge 0$ ensure the positivity of $y_t$.

The way we model $\mathbf{U}_t$ and $\mathbf{V}_t$ in (\ref{Matrix_volatility_U})--(\ref{Matrix_volatility_V}) is new to the literature, and it is
well suited to the matrix nature of $\mathbf{X}_t$.
In (\ref{MGARCH_1})--(\ref{MGARCH_2}), both $\mathbf{S}_{1,t}$ and $\mathbf{S}_{2,t}$ have the BEKK specifications (\citealp{engle1995multivariate}) to capture the dynamics of conditional row and column covariance matrices, respectively. However, due to the aforementioned identification issue, $\mathbf{U}_t$ and $\mathbf{V}_t$ are not directly related to $\mathbf{S}_{1,t}$ and $\mathbf{S}_{2,t}$ directly, but rather to their normalized versions $\mathbf{S}_{1,t}/\text{tr}(\mathbf{S}_{1,t})$ and $\mathbf{S}_{2,t}/\text{tr}(\mathbf{S}_{2,t})$.
The normalization of $\mathbf{S}_{2,t}$ in (\ref{Matrix_volatility_V}) guarantees $\text{tr}(\mathbf{V}_t)=1$, so that the identification problem can be solved. The normalization of $\mathbf{S}_{1,t}$ in (\ref{Matrix_volatility_U}) ensures that $\text{tr}(\mathbf{U}_t)=y_t$, and together with the condition $\text{tr}(\mathbf{V}_t)=1$, we have
\begin{flalign*}
\text{tr}\left(E\left(\mathbf{X}_t\mathbf{X}_t'|\mathcal{F}_{t-1}\right)\right)
=\text{tr}\left(E\left(\mathbf{X}_t'\mathbf{X}_t|\mathcal{F}_{t-1}\right)\right)
=\text{tr}(\mathbf{V}_t)\text{tr}(\mathbf{U}_t)=y_t.
\end{flalign*}
The above result implies that under (\ref{Matrix_volatility_U})--(\ref{Matrix_volatility_V}), $y_t$ is indeed the trace of conditional row or column covariance matrix of $\mathbf{X}_t$,
which motivates its univariate GARCH specification (\citealp{bollerslev1986generalized}) in (\ref{MGARCH_3}).
Moreover, since the specifications of $\mathbf{U}_t$ and $\mathbf{V}_t$ are invariant to scale changes of $\mathbf{S}_{1,t}$ and $\mathbf{S}_{2,t}$, we add the constraints in (\ref{A_0_B_0}) to avoid this
ambiguity.

In the sequel, we call model (\ref{Eq2.2}) with the specifications of $\mathbf{U}_t$ and $\mathbf{V}_t$ in  (\ref{Matrix_volatility_U})--(\ref{Matrix_volatility_V}) the matrix GARCH model.
By (\ref{row_cov})--(\ref{column_cov}), we know that the conditional row and column covariance matrices of $\mathbf{X}_t$
under the matrix GARCH model are $\mathbf{U}_t$ and $y_t\mathbf{V}_t$, respectively.
Intrinsically, the matrix GARCH model is related to univariate or multivariate GARCH-type models.
For example, when $m=n=1$, we have $\mathbf{U}_{t} = y_t$, $\mathbf{V}_{t}= 1$, and $\mathbf{X}_t = y_t\mathbf{Z}_t$, so the matrix GARCH model reduces to the univariate GARCH model in \cite{bollerslev1986generalized}.
When $n=1$, $m>1$, and $y_t/\text{tr}(\mathbf{S}_{1,t}) = c$ for some constant $c>0$, we have $\mathbf{U}_t=c\mathbf{S}_{1,t}$, $\mathbf{V}_t=1$, and $\mathbf{X}_t=c\mathbf{S}_{1,t}\mathbf{Z}_t$, so the matrix GARCH model becomes the BEKK model in \cite{engle1995multivariate}.
Note that the condition  $y_t/\text{tr}(\mathbf{S}_{1,t}) = c$ holds, when
$\mathbf{A}_1 = a_1\mathbf{I}_m$, $\mathbf{A}_2 = a_2\mathbf{I}_m$, $w = c\|\mathbf{A}_0 \|^2$, $\alpha = c(a_1m)^2$, and $\beta = c(a_2m)^2$.

Using the vectorization operator, the matrix GARCH model can also be represented as
\begin{equation}\label{vec_garch}
\text{vec}(\mathbf{X}_t) = \mathbf{\Sigma}_t^{1/2} \text{vec}(\mathbf{Z}_t),
\end{equation}
where $\mathbf{\Sigma}_t$ is the $mn \times mn$ conditional covariance matrix of $\text{vec}(\mathbf{X}_t)$ with the Kronecker product structure
\begin{equation}\label{Vector}
\mathbf{\Sigma}_t = \mathbf{V}_t \otimes \mathbf{U}_t = \frac{y_t}{\text{tr}(\mathbf{S}_{1,t}) \operatorname{tr}(\mathbf{S}_{2,t})} \mathbf{S}_{2,t} \otimes \mathbf{S}_{1,t}.
\end{equation}
However, from (\ref{vec_garch})--(\ref{Vector}), we see that except for some special cases (as demonstrated above), the matrix GARCH model in general  does not have the form of the BEKK model,
for which $\mathbf{\Sigma}_t$ has the specification
\begin{align}\label{bekk_model}
\mathbf{\Sigma}_t = \mathbf{C}_0\mathbf{C}_0' + \mathbf{C}_1\text{vec}(\mathbf{X}_{t-1}) \text{vec}(\mathbf{X}_{t-1})'\mathbf{C}_1' + \mathbf{C}_2\mathbf{\Sigma}_{t-1}\mathbf{C}_2',
\end{align}
where $\mathbf{C}_0$, $\mathbf{C}_1$, and $\mathbf{C}_2$ are $mn\times mn$ parameter matrices defined similarly as those in (\ref{MGARCH_1}). One may use the BEKK model in (\ref{bekk_model}) (or other multivariate GARCH-type models) to model $\text{vec}(\mathbf{X}_t)$ directly. However, this approach has two disadvantages.
First, the matrix structure of $\mathbf{X}_t$ is brutally broken, so the fitting results from the BEKK model
lack interpretations on conditional row or column covariance. Second, the number of parameters in the BEKK model has an order of $O(m^2n^2)$, whereas that in the matrix GARCH model has a much smaller order of $O(m^2+n^2)$, even for the moderate values of $m$ and $n$.

Needless to say,  the matrix GARCH model can be easily extended
to the case where $\mathbf{S}_{1,t}$ and $\mathbf{S}_{2,t}$ have   higher order BEKK specifications,
and $y_t$ has a higher order GARCH specification.
With slight modifications, our proposed methodologies and their related technical treatments below are available
for this extended higher order model, which is thus not investigated in details for ease of exposition.

\section{Quasi Maximum Likelihood Estimation}
\label{sec.estimation}

This section studies the QMLE of our matrix GARCH model. Let $\theta = (\gamma', \delta', \zeta')' \in \Theta$  be a vector of unknown parameters with the true value $\theta_0$
for the matrix GARCH model, where $\gamma=(w, \alpha, \beta)'\in\Theta_{\gamma}\subset \mathbb{R}^{3}$, $\delta=(\text{vech}(\mathbf{A}_0)',\text{vec}(\mathbf{A}_1)',\text{vec}(\mathbf{A}_2)')'\in\Theta_{\delta}\subset \mathbb{R}^{p_{\delta}}$ with
$p_{\delta}=2m^2+[m(m+1)]/2-1$,
$\zeta=(\text{vech}(\mathbf{B}_0)', \text{vec}(\mathbf{B}_1)', \text{vec}(\mathbf{B}_2)')'\in\Theta_{\zeta}\subset \mathbb{R}^{p_{\zeta}}$ with
$p_{\zeta}=2n^2+[n(n+1)]/2-1$,
$\Theta=\Theta_{\gamma}\times\Theta_{\delta}\times\Theta_{\zeta}\subset\mathbb{R}^{p}$ is a compact parameter space with $p=3+p_{\delta}+p_{\zeta}$, and $\theta_0\in\Theta$.
Here, due to the constraints in (\ref{A_0_B_0}), we exclude
$\mathbf{A}_{0,11}$ and $\mathbf{B}_{0,11}$ in  $\text{vech}(\mathbf{A}_0)$ and $\text{vech}(\mathbf{B}_0)$, respectively.


By assuming that $\mathbf{Z}_t$ follows the standard matrix normal distribution $\text{MN}(\mathbf{0}, \mathbf{I}_m, \mathbf{I}_n)$ (see its definition in Chapter 2 of \cite{gupta1999matrix}), we have $\text{vec}(\mathbf{X}_t) | \mathcal{F}_{t-1} \sim \text{N}(\mathbf{0}, \mathbf{\Sigma}_t)$ with $\mathbf{\Sigma}_t = \mathbf{V}_t \otimes \mathbf{U}_t$. Then the quasi log-likelihood function (ignoring constants) of $\{\text{vec}(\mathbf{X}_t)\}_{t=1}^{T}$ is
 \begin{align*}
L_T(\theta) &= \frac{1}{T}\sum_{t=1}^{T}l_t(\theta):=\frac{1}{2T}\sum_{t=1}^{T}\left\{\log|\mathbf{\Sigma}_t(\theta)| +  \text{vec}(\mathbf{X}_t)'\mathbf{\Sigma}_t^{-1}(\theta) \text{vec}(\mathbf{X}_t)\right\},
\end{align*}
where
$$\mathbf{\Sigma}_t(\theta)= \frac{y_t(\gamma)}{\text{tr}(\mathbf{S}_{1,t}(\delta))\text{tr}(\mathbf{S}_{2,t}(\zeta))} \mathbf{S}_{2,t}(\zeta) \otimes \mathbf{S}_{1,t}(\delta)$$
with $\mathbf{S}_{1,t}(\delta)$, $\mathbf{S}_{2,t}(\zeta)$, and $y_t(\gamma)$ being calculated recursively by
\begin{align*}
\mathbf{S}_{1,t}(\delta) &= \mathbf{A}_0 \mathbf{A}_0' + \mathbf{A}_1\mathbf{X}_{t-1} \mathbf{X}_{t-1}' \mathbf{A}_1' + \mathbf{A}_2 \mathbf{S}_{1,t-1}(\delta) \mathbf{A}_2', \\
\mathbf{S}_{2,t}(\zeta) &= \mathbf{B}_0 \mathbf{B}_0' + \mathbf{B}_1\mathbf{X}_{t-1}' \mathbf{X}_{t-1} \mathbf{B}_1' + \mathbf{B}_2 \mathbf{S}_{2,t-1}(\zeta) \mathbf{B}_2', \\
y_t(\gamma) &= w + \alpha\, \text{tr}(\mathbf{X}_{t-1} \mathbf{X}_{t-1}') + \beta y_{t-1}(\gamma).
\end{align*}
To compute $L_T(\theta)$, the initial values $\{ \mathbf{X}_t:  t \leq 0\}$ are necessary, which are unfortunately unobservable. Thus, we consider the computationally feasible version of $L_T(\theta)$ as follows
\begin{equation*}
\widetilde{L}_T(\theta) = \frac{1}{T}\sum_{t=1}^{T}\widetilde{l}_t(\theta),
\end{equation*}
where $\widetilde{l}_t(\theta)$ is defined in the same way as $l_t(\theta)$ with $\mathbf{\Sigma}_t(\theta)$ being replaced by $\widetilde{\mathbf{\Sigma}}_t(\theta)$, and  $\widetilde{\mathbf{\Sigma}}_t(\theta)$ is calculated in the same way as  $\mathbf{\Sigma}_t(\theta)$ based on the given initial values $\mathbf{X}_0=\mathbf{0}$, $\mathbf{S}_{1,0}=\mathbf{0}$, $\mathbf{S}_{2,0}=\mathbf{0}$, and $y_{0}=0$.

Let $\widehat{\theta}$ be the minimizer of $\widetilde{L}_T(\theta)$, that is,
\begin{equation*} 
\widehat{\theta} = \mathop{\arg\min}\limits_{\theta \in \Theta} \widetilde{L}_T(\theta).
\end{equation*}
Although $\widehat{\theta}$ is defined by assuming $\mathbf{Z}_t\sim\text{MN}(\mathbf{0}, \mathbf{I}_m, \mathbf{I}_n)$, the consistency and asymptotic normality of $\widehat{\theta}$ below still hold for general distributions of $\mathbf{Z}_t$ with some moment constraints. Thus, we call $\widehat{\theta}$ the QMLE of $\theta_0$.


To study the asymptotic properties of $\widehat{\theta}$, some technical assumptions are needed.

\begin{assumption}\label{ass_stationary}
	$\{\mathbf{X}_t\}$ is strictly stationary and ergodic.
\end{assumption}

\begin{assumption}\label{ass_identification}
If $\mathbf{\Sigma}_t(\theta)=\mathbf{\Sigma}_t(\theta_0)$ almost surely for any $\theta\in\Theta$, then
	$\theta=\theta_0$.
\end{assumption}

\begin{assumption}\label{ass_parameter}
There exist some constants $\underline{\rho}\in (0,1)$ and $s>0$ such that

\noindent $(i)$.  $\rho(\mathbf{A}_1\otimes \mathbf{A}_1+\mathbf{A}_2\otimes \mathbf{A}_2)\leq \underline{\rho}$, $\rho(\mathbf{B}_1\otimes \mathbf{B}_1+\mathbf{B}_2\otimes \mathbf{B}_2)\leq \underline{\rho}$, and $\alpha+\beta \leq \underline{\rho}$;

\noindent $(ii)$.  $\displaystyle\sup _{\theta \in \Theta} \min\{\|\mathbf{A}_2\|^2,\|\mathbf{B}_2\|^2\}\beta^{s-1} \le \underline{\rho}$ and  $\displaystyle\sup _{\theta \in \Theta} \max\{\|\mathbf{A}_2\|^2,\|\mathbf{B}_2\|^2\}^{s-1}\beta \le \underline{\rho}$.
\end{assumption}

Assumption \ref{ass_stationary} is commonly used for statistical inference on nonlinear time series models. For multivariate GARCH-type models, the conditions for their stationarity and ergodicity are obtained by using the stochastic recurrence equation theory (\citealp{Ling2003asymptotic,Hafner2009asymptotic,boussama2011stationarity}). However, such an approach seems inapplicable for the matrix GARCH model, since its $\mathbf{\Sigma}_t$ does not have a tractable recursive form.
Assumption \ref{ass_identification} is the standard identification condition originated from multivariate GARCH-type models; see \cite{comte2003asymptotic}, \cite{Hafner2009asymptotic}, and \cite{Zhou2021CBF}.
Assumption \ref{ass_parameter} gives some technical conditions to bound $E\|\mathbf{\Sigma}_t^{-1}(\theta) \|$ and is used to deal with the initial value problem. Assumption \ref{ass_parameter}(i) is similar to sufficient conditions for the covariance stationarity of
the BEKK or univariate GARCH models, and
Assumption \ref{ass_parameter}(ii) is mild as it holds trivially when $s\rightarrow 1$.

Now, we are ready to establish the consistency and asymptotic normality of $\widehat{\theta}$.

\begin{theorem}\label{TH1}
If Assumptions \ref{ass_stationary}--\ref{ass_parameter} hold
and $\text{E}\|\mathbf{X}_t\|^{8+s} < \infty$, then, $\widehat{\theta} \stackrel{p}{\to} \theta_0$
as $T \rightarrow \infty$.
\end{theorem}

\begin{theorem}\label{TH2}
If Assumptions \ref{ass_stationary}--\ref{ass_parameter} hold, $\theta_0$ is an interior point of $\Theta$, $\text{E}\|\mathbf{X}_t\|^{6(2+s)} < \infty$, $\mathcal{C}_0=E\left(\frac{\partial^{2} l_{t}\left(\theta_{0}\right)}{\partial \theta \partial \theta^{\prime}}\right)$ is invertible, and $\mathcal{C}_1 = E\left(\frac{\partial l_t(\theta_0)}{\partial \theta} \frac{\partial l_t(\theta_0)}{\partial \theta'}   \right)$ is positive definite, then, $\sqrt{T}\big(\widehat{\theta}-\theta_{0}\big) \stackrel{d}{\rightarrow} \text{N}\left(0, \mathcal{C}_0^{-1} \mathcal{C}_1 \mathcal{C}_0^{-1}  \right) \text { as } T \rightarrow \infty$.
\end{theorem}


In the two aforementioned theorems, the moment conditions of $\mathbf{X}_t$ are deemed sufficient. However, we are of the opinion that a more intricate analysis could potentially aid in reducing these requirements.
Using $\widehat{\theta}$, we can estimate $\mathcal{C}_0$ and $\mathcal{C}_1$ in Theorem \ref{TH2} by
\begin{align*}
\widehat{\mathcal{C}}_0=\frac{1}{T}\sum_{t=1}^{T}\frac{\partial^{2} \widetilde{l}_{t}(\widehat{\theta})}{\partial \theta \partial \theta^{\prime}}
\,\,\,\mbox{ and }\,\,\,
\widehat{\mathcal{C}}_1=\frac{1}{T}\sum_{t=1}^{T}\frac{\partial \widetilde{l}_t(\widehat{\theta})}{\partial \theta} \frac{\partial \widetilde{l}_t(\widehat{\theta})}{\partial \theta'},
\end{align*}
respectively. Under the conditions of Theorem \ref{TH2}, it is not hard to see that $\widehat{\mathcal{C}}_0$ and $\widehat{\mathcal{C}}_1$ are consistent estimators of $\mathcal{C}_0$ and $\mathcal{C}_1$, respectively. Based on  $\widehat{\mathcal{C}}_0$ and $\widehat{\mathcal{C}}_1$, the asymptotic standard errors of $\widehat{\theta}$ can be easily computed.

\section{Model Diagnostic Checking}
\label{sec: Model Checking}

Diagnostic checking is a crucial step in time series analysis. For multivariate GARCH-type models, several diagnostic checking methods have been proposed in the literature; see, for example, \cite{ling1997diagnostic}, \cite{tse2002residual}, and \cite{escanciano2013automatic}.
In this section, we follow the idea of \cite{ling1997diagnostic} and then propose a new portmanteau test for checking the adequacy of a fitted matrix GARCH model.

Let $\widehat{\mathbf{Z}}_t = \widehat{\mathbf{U}}_t^{-1/2} \mathbf{X}_t\widehat{\mathbf{V}}_t^{-1/2}$ be the residual of
a fitted matrix GARCH model, where $\widehat{\mathbf{U}}_t$ and $\widehat{\mathbf{V}}_t$ are computed based on the QMLE $\widehat{\theta}$.
Note that if the matrix GARCH model is correctly specified, $\{\text{vec}(\mathbf{Z}_t)\}_{t=1}^{T}$ is a sequence of i.i.d. random vectors with $E\left[\text{vec}(\mathbf{Z}_t) \text{vec}(\mathbf{Z}_t)'\right] = \mathbf{I}_{mn}$. In this case, it turns out that
 $\{\text{vec}(\mathbf{Z}_t)'\text{vec}(\mathbf{Z}_t)-mn\}_{t=1}^{T}$ is a sequence of uncorrelated random variables.
 Hence, if the matrix GARCH model is correctly specified,
it is expected that the sample autocorrelation of $\{\text{vec}(\widehat{\mathbf{Z}}_t)'\text{vec}(\widehat{\mathbf{Z}}_t)-mn\}_{t=1}^{T}$ at lag $l(\neq 0)$, denoted as $\widehat{R}_l$, is close to zero, where
\begin{equation}\label{Auto_residual}
	\widehat{R}_l = \frac{\sum_{t=l+1}^{T}\left(\text{vec}(\mathbf{X}_t)^{'} \widehat{\mathbf{\Sigma}}_t^{-1}\text{vec}(\mathbf{X}_t) - mn \right)\left(\text{vec}(\mathbf{X}_{t-l})^{'} \widehat{\mathbf{\Sigma}}_{t-l}^{-1}\text{vec}(\mathbf{X}_{t-l}) - mn \right)}{\sum_{t=1}^{T}\left(\text{vec}(\mathbf{X}_t)^{'} \widehat{\mathbf{\Sigma}}_t^{-1}\text{vec}(\mathbf{X}_t) - mn \right)^2}
\end{equation}
with $\widehat{\mathbf{\Sigma}}_t=\widehat{\mathbf{V}}_t\otimes\widehat{\mathbf{U}}_t$.

For  a given integer $L \ge 1$, let $\widehat{R} = (\widehat{R}_1, \widehat{R}_2, ..., \widehat{R}_L)'\in\mathbb{R}^{L}$,
$\mathbf{M} = (M_1, M_2, ..., M_L)'\in\mathbb{R}^{L\times p}$ and $\mathbf{N} = (N_1, N_2, ..., N_L)'\in\mathbb{R}^{L \times p}$, where
\begin{align*}
M_l =& E\left[ \frac{\partial \text{vec}(\mathbf{\Sigma}_t(\theta_0))}{\partial \theta' }\text{vec}\Big(\mathbf{\Sigma}_t(\theta_0)^{-1}\big\{\text{vec}(\mathbf{X}_{t-l})' \mathbf{\Sigma}_{t-l}(\theta_0)^{-1}\text{vec}(\mathbf{X}_{t-l}) - mn\big\}\Big)\right],\\
N_l =& E\left[ \frac{\partial \text{vec}(\mathbf{\Sigma}_t(\theta_0))}{\partial \theta' }\text{vec}\Big(\mathbf{\Sigma}_t(\theta_0)^{-1} \text{vec}(\mathbf{X}_t) \text{vec}(\mathbf{X}_t) '  \mathbf{\Sigma}_t(\theta_0)^{-1}\Big)\right.\\
&\quad\left.\times\Big(\text{vec}(\mathbf{X}_{t})' \mathbf{\Sigma}_{t}(\theta_0)^{-1}\text{vec}(\mathbf{X}_{t}) - mn\Big)\Big(\text{vec}(\mathbf{X}_{t-l})' \mathbf{\Sigma}_{t-l}(\theta_0)^{-1}\text{vec}(\mathbf{X}_{t-l}) - mn\Big)\right].
\end{align*}
Moreover, let $\kappa = E\big( \sum_{i =1}^{m}\sum_{j=1}^{n}\mathbf{Z}_{t,ij}^2  \big)^2 - m^2n^2$ and $\boldsymbol{\eta}=(\eta_{11},\eta_{12},...,\eta_{mn})'\in\mathbb{R}^{mn}$ with $\eta_{ij} = E(\mathbf{Z}_{t,ij}^4) - 1$.
To facilitate our portmanteau test, we need the limiting distribution of $\widehat{R}$ below.

\begin{theorem}\label{TH3}
Suppose that the conditions in Theorem \ref{TH2} hold and $E\|\mathbf{Z}_{t}\|^4 < \infty$. If the matrix GARCH model  is correctly specified, then
	\begin{equation*}
		\sqrt{T} \widehat{R} \stackrel{d}{\rightarrow} \text{N}(\mathbf{0}, \mathbf{\Omega}) \quad \text{as } T \rightarrow \infty,
	\end{equation*}
	where  $\mathbf{\Omega} =  \left([\mathbf{1}_{mn}'\boldsymbol{\eta}]^2\mathbf{I}_{L} -\left( \mathbf{N}\mathcal{C}_0^{-1} \mathbf{M}'/2 + \mathbf{M}\mathcal{C}_0^{-1} \mathbf{N}'/2 -  \mathbf{M}\mathcal{C}_0^{-1}\mathcal{C}_1\mathcal{C}_0^{-1} \mathbf{M}'\right) \right)  / \kappa^2$.
\end{theorem}

Based on Theorem \ref{TH3}, we construct the portmanteau test statistic at the lag $L$ as
\begin{equation*} 
	Q_T(L) = T \big[ \widehat{R}'\widehat{\mathbf{\Omega}}^{-1}\widehat{R}\big],
\end{equation*}
where $\widehat{\mathbf{\Omega}}$ is the sample counterpart of $\mathbf{\Omega}$. If the matrix GARCH model is correctly specified, then $Q_T(L) \stackrel{d}{\rightarrow} \chi_{L}^2$ as $T \rightarrow \infty$. Thus, when the value of $Q_T(L)$ is larger than the $\alpha^*$-th upper-tailed critical value of $\chi_{L}^2$, the fitted matrix GARCH model is inadequate at the significance level $\alpha^*$. Otherwise, it is adequate.
In practice, the choice of $L$ depends
on the frequency of the series, and one can often choose $L$ to be
$O(\log(T))$, which delivers 6 or 8 for a moderate value of $T$ (see \cite{tsay2008}).

\section{Matrix Factor GARCH Model}
\label{sec: Matrix Factor GARCH}


Considering that the number of parameters in the matrix GARCH model has an order of $O(m^2 + n^2)$, the estimation would be  computationally challenging when the dimension $m$ or $n$ is large.  To deal with the matrix data with large dimension, we adopt the idea of matrix factor model (\citealp{Wang2019Factor}) by assuming that $\mathbf{X}_t$ is driven by a $k_1 \times k_2$ latent common factor matrix $\mathbf{F}_t$, with $k_1$ and $k_2$ much smaller than $m$ and $n$. In particular, we assume
\begin{equation}\label{MatirxFactor}
\mathbf{X}_t = \mathbf{R}^*  \mathbf{F}_t^* (\mathbf{C}^*)' + \mathbf{E}_t,
\end{equation}
where $\mathbf{R}^*$ is an $m \times k_1$ row factor loading matrix exploiting the variations of $\mathbf{X}_t$ across the rows, $\mathbf{C}^*$ is an $n \times k_2$ column factor loading matrix reflecting the variations of $\mathbf{X}_t$ across the columns, $\mathbf{E}_t$ is an $m\times n$ idiosyncratic matrix white noise error with $E(\mathbf{E}_t)=0$ and
 $\mathbf{\Sigma}_e = E\left[\text{vec}(\mathbf{E}_t) \text{vec}(\mathbf{E}_t)' \right]<\infty$, and $\mathbf{E}_t$ is also uncorrelated with $\mathbf{F}_{t}^*$.
For simplicity, we assume that
$(\mathbf{R}^*)'\mathbf{R}^*=\mathbf{I}_{k_1}$ and $(\mathbf{C}^*)'\mathbf{C}^*=\mathbf{I}_{k_2}$.


For model (\ref{MatirxFactor}),  \cite{YU2022projected} propose the projection estimators $\widehat{\mathbf{R}}^*$ and $\widehat{\mathbf{C}}^*$, and under some regular conditions,
they show that
\begin{equation}\label{projection_est}
\|\widehat{\mathbf{R}}^*-\mathbf{R}^*\mathbf{H}_{1}\| \stackrel{p}{\to}0 \mbox{ and }\|\widehat{\mathbf{C}}^*-\mathbf{C}^*\mathbf{H}_{2}\| \stackrel{p}{\to} 0 \mbox{ as } T, m, n\to\infty
\end{equation}
for two orthogonal matrices $\mathbf{H}_{1}\in\mathcal{R}^{k_1\times k_1}$ and $\mathbf{H}_{2}\in\mathcal{R}^{k_2\times k_2}$. See also
\cite{Wang2019Factor} and \cite{chen2021statistical} for other estimators of loading matrices.
Due to the factor nature of model (\ref{MatirxFactor}), instead of considering the estimated loading matrices $\widehat{\mathbf{R}}^*$ and $\widehat{\mathbf{C}}^*$  directly,  we follow \cite{chen2022factor} to consider
their varimax rotated versions (\citealp{kaiser1958varimax}) given by $\widehat{\mathbf{R}}^*\mathbf{G}_1$ and $\widehat{\mathbf{C}}^*\mathbf{G}_2$, where
$\mathbf{G}_1$ and $\mathbf{G}_2$ are $k_1\times k_1$ and $k_2\times k_2$ orthogonal matrices, respectively.
We focus on the varimax rotated loading matrices, because varimax rotation brings  simpler structure so that each factor will have fewer variables with large loadings, leading to a better interpretation of factors.

The consistency results in (\ref{projection_est}) and varimax rotation motivate us to reshape model (\ref{MatirxFactor}) as
\begin{equation}\label{MatirxFactor_1}
\mathbf{X}_t = \mathbf{R}  \mathbf{F}_t \mathbf{C}' + \mathbf{E}_t,
\end{equation}
where $\mathbf{R}=\mathbf{R}^*\mathbf{H}_{1}\mathbf{G}_1$, $\mathbf{C}=\mathbf{C}^*\mathbf{H}_{2}\mathbf{G}_2$, and $\mathbf{F}_t=\mathbf{G}_1'\mathbf{H}_{1}'\mathbf{F}_t^*\mathbf{H}_{2}'\mathbf{G}_2'$ that is assumed to follow the matrix GARCH model:
$\mathbf{F}_t=\mathbf{U}_{f,t}^{1/2}\mathbf{Z}_t\mathbf{V}_{f,t}^{1/2}$. Here,
$\mathbf{U}_{f,t}$ and $\mathbf{V}_{f,t}$ are defined in the same way as $\mathbf{U}_{t}$ and $\mathbf{V}_{t}$ in (\ref{Eq2.2}) with $\mathbf{X}_t$ replaced by $\mathbf{F}_t$, and $\mathbf{Z}_t$ is independent of $\mathcal{F}_{t-1}$.
We call model (\ref{MatirxFactor_1}) the matrix factor GARCH model, since $\mathbf{F}_t$ has the matrix GARCH structure.

Using the matrix factor GARCH model, we are able to study the high-dimensional conditional covariance matrix of
$\text{vec}(\mathbf{X}_t)$, which is useful for portfolio selections based on many assets that have a matrix time series structure in nature (see our two applications in Section \ref{sec.real} below). Specifically, we have $\text{vec}(\mathbf{X}_t) = (\mathbf{C}\otimes \mathbf{R})\text{vec}(\mathbf{F}_t) + \text{vec}(\mathbf{E}_t)$ under (\ref{MatirxFactor_1}), so the conditional covariance matrix of $\text{vec}(\mathbf{X}_t)$ is
\begin{equation*}
\mathbf{\Sigma}_{x,t} := E\left[\text{vec}(\mathbf{X}_t)\text{vec}(\mathbf{X}_t)'|\mathcal{F}_{t-1}\right]
=(\mathbf{C}\otimes \mathbf{R}) \mathbf{\Sigma}_{f,t} (\mathbf{C}\otimes \mathbf{R})' + \mathbf{\Sigma}_{e},
\end{equation*}
where $\mathbf{\Sigma}_{f,t}=\mathbf{V}_{f,t}\otimes \mathbf{U}_{f,t}$.
Let $\widehat{\mathbf{R}}^{**}=\widehat{\mathbf{R}}^*\mathbf{G}_1$, $\widehat{\mathbf{C}}^{**}=\widehat{\mathbf{C}}^*\mathbf{G}_2$, and
$\widehat{\mathbf{F}}_t = (\widehat{\mathbf{R}}^{**})' \mathbf{X}_t \widehat{\mathbf{C}}^{**}$.
In view of (\ref{projection_est}), we can estimate $\mathbf{\Sigma}_{x,t}$ by
\begin{equation}\label{est_sigma_x}
\widehat{\mathbf{\Sigma}}_{x,t}=(\widehat{\mathbf{C}}^{**}\otimes \widehat{\mathbf{R}}^{**}) (\widehat{\mathbf{V}}_{f,t}\otimes \widehat{\mathbf{U}}_{f,t}) (\widehat{\mathbf{C}}^{**}\otimes \widehat{\mathbf{R}}^{**})'+\widehat{\mathbf{\Sigma}}_e,
\end{equation}
where $\widehat{\mathbf{V}}_{f,t}$ and $\widehat{\mathbf{U}}_{f,t}$ are estimators of
$\mathbf{V}_{f,t}$ and $\mathbf{U}_{f,t}$ computed by using the QMLE based on $\{\widehat{\mathbf{F}}_t\}_{t=1}^{T}$,
$\widehat{\mathbf{\Sigma}}_e = T^{-1}\sum_{t=1}^{T}\text{vec}(\widehat{\mathbf{E}}_t)\text{vec}(\widehat{\mathbf{E}}_t)'$ with $\widehat{\mathbf{E}}_t = \mathbf{X}_t - \widehat{\mathbf{R}}^{**} \widehat{\mathbf{F}}_t (\widehat{\mathbf{C}}^{**})'$, and
the number of factors $k_1$ and $k_2$ is determined by the eigenvalue ratio method in \cite{YU2022projected}.

\section{Simulations}\label{sec.sim}
\subsection{Simulation for the QMLE}
In this subsection, we assess the finite sample performance of the QMLE $\widehat{\theta}$. We generate $1000$ replications of sample size $T=1000$ and $2000$ from the following matrix GARCH model
\begin{equation*}
    \mathbf{X}_t = \mathbf{U}_t^{1/2} \mathbf{Z}_t \mathbf{V}_t^{1/2} ,
\end{equation*}
where
\begin{align*}
    \mathbf{U}_t &= \frac{ \mathbf{S}_{1t} }{ \text{tr}(\mathbf{S}_{1t} )  } y_t, \,\,\,\,\mathbf{V}_t = \frac{ \mathbf{S}_{2t} }{ \text{tr}(\mathbf{S}_{2t} ) },\,\,\,\,\mathbf{S}_{1t} = \mathbf{A}_0 \mathbf{A}_0' + \mathbf{A}_1\mathbf{X}_{t-1} \mathbf{X}_{t-1}' \mathbf{A}_1' + \mathbf{A}_2 \mathbf{S}_{1,t-1} \mathbf{A}_2', \\
\mathbf{S}_{2t} &= \mathbf{B}_0 \mathbf{B}_0' + \mathbf{B}_1\mathbf{X}_{t-1}^{'} \mathbf{X}_{t-1} \mathbf{B}_1' + \mathbf{B}_2 \mathbf{S}_{2,t-1} \mathbf{B}_2', \,\,\,\,
y_t = w + \alpha \text{tr}(\mathbf{X}_{t-1} \mathbf{X}_{t-1}') + \beta y_{t-1}
\end{align*}
with the parameters $w =0.4, \alpha = 0.3, \beta = 0.6$,
\begin{align*}
\mathbf{A}_0 &=\mathbf{B}_0 = \begin{pmatrix} 1& 0 &0 \\
0.4 & 0.4 & 0\\
0.4 & 0.4 & 0.4 \end{pmatrix},
\mathbf{A}_1 = \mathbf{B}_1 =\begin{pmatrix} 0.3 & 0 & 0 \\ 0 & 0.3 &0 \\
0 & 0& 0.3\end{pmatrix},
\mathbf{A}_2 = \mathbf{B}_2 =\begin{pmatrix} 0.6 & 0 & 0 \\ 0 & 0.6 &0 \\
0 & 0& 0.6\ \end{pmatrix},
\end{align*}%
and $\{\mathbf{Z}_t\}_{t=1}^{T}$ is an i.i.d. sequence with $\mathbf{Z}_t\sim \text{MN}(\mathbf{0}, \mathbf{I}_3, \mathbf{I}_3)$, $\text{SMT}_{15}$, and $\text{SMT}_{25}$. Here, $\text{SMT}_{\nu}$ is the standardized matrix $t$ distribution with degrees of freedom $\nu$, row covariance $\mathbf{I}_3$, and column covariance $\mathbf{I}_3$.

Based on 1000 replications, Tables \ref{Table_Normal}, \ref{Table_T15}, and \ref{Table_T25} report the bias,
sample root mean squared error (SE), and averaged asymptotic standard error (AE) of $\widehat{\theta}$ when $\mathbf{Z}_t$ follows $\text{MN}(\mathbf{0}, \mathbf{I}_3, \mathbf{I}_3)$, $\text{SMT}_{15}$, and $\text{SMT}_{25}$ distributions, respectively. From these tables, we find that (i) regardless of the distribution of $\mathbf{Z}_t$, $\widehat{\theta}$ has small biases, and its
values of SE are close to those of AE; (ii) the values of bias and AE decrease as the sample size $T$ increases; (iii)
as expected, the value of AE in the case of $\mathbf{Z}_t\sim \text{MN}(\mathbf{0}, \mathbf{I}_3, \mathbf{I}_3)$ is smaller than that in the case of $\mathbf{Z}_t\sim \text{SMT}_{15}$ or $\text{SMT}_{25}$. Overall, our proposed QMLE $\widehat{\theta}$ has a good finite-sample performance in all examined cases.


\begin{table}[htbp]
  \scriptsize
  \centering
  \caption{The results of $\widehat{\theta}$ when $\mathbf{Z}_t\sim \text{MN}(\mathbf{0}, \mathbf{I}_3, \mathbf{I}_3)$.}
    \begin{tabular}{clccccccccccccc}
    \toprule
        $T$  &       & $w$     & $\alpha$ & $\beta$  & $\mathbf{A}_{0,21}$ & $\mathbf{A}_{0,22}$ & $\mathbf{A}_{0,31}$ & $\mathbf{A}_{0,32}$ & $\mathbf{A}_{0,33}$ & $\mathbf{A}_{1,11}$ & $\mathbf{A}_{1,22}$ & $\mathbf{A}_{1,33}$ & $\mathbf{A}_{2,11}$ & $\mathbf{A}_{2,22}$\\
    \midrule
    1000 & Bias  & 0.021  & -0.003  & -0.007  & 0.017  & 0.023  & -0.010  & 0.003  & -0.015  & 0.007  & 0.005  & 0.006  & -0.020  & -0.010  \\
          & SE   & 0.052  & 0.024  & 0.032  & 0.067  & 0.081  & 0.056  & 0.071  & 0.075  & 0.040  & 0.050  & 0.055  & 0.144  & 0.122  \\
          & AE   & 0.049  & 0.023  & 0.031  & 0.061  & 0.072  & 0.053  & 0.068  & 0.070  & 0.037  & 0.046  & 0.051  & 0.132  & 0.116  \\
\cmidrule{2-15}          &       & $\mathbf{A}_{2,33}$ & $\mathbf{B}_{0,21}$ & $\mathbf{B}_{0,22}$ & $\mathbf{B}_{0,31}$ & $\mathbf{B}_{0,32}$ & $\mathbf{B}_{0,33}$ & $\mathbf{B}_{1,11}$ & $\mathbf{B}_{1,22}$ & $\mathbf{B}_{1,33}$ & $\mathbf{B}_{2,11}$ & $\mathbf{B}_{2,22}$ & $\mathbf{B}_{2,33}$ &  \\
\cmidrule{2-15}          & Bias  & -0.014  & 0.022  & 0.027  & -0.007  & 0.004  & -0.013  & 0.007  & 0.008  & 0.010  & -0.013  & -0.016  & -0.017  &  \\
          & SE   & 0.134  & 0.074  & 0.091  & 0.058  & 0.074  & 0.078  & 0.039  & 0.049  & 0.054  & 0.141  & 0.122  & 0.139  &  \\
          & AE   & 0.125  & 0.070  & 0.080  & 0.055  & 0.073  & 0.076  & 0.037  & 0.041  & 0.050  & 0.129  & 0.116  & 0.125  &  \\
    \midrule
         $T$ &       & $w$     & $\alpha$ & $\beta$  & $\mathbf{A}_{0,21}$ & $\mathbf{A}_{0,22}$ & $\mathbf{A}_{0,31}$ & $\mathbf{A}_{0,32}$ & $\mathbf{A}_{0,33}$ & $\mathbf{A}_{1,11}$ & $\mathbf{A}_{1,22}$ & $\mathbf{A}_{1,33}$ & $\mathbf{A}_{2,11}$ & $\mathbf{A}_{2,22}$\\
    \midrule
    2000 & Bias  & 0.010  & 0.000  & -0.004  & 0.007  & 0.010  & -0.004  & 0.001  & -0.006  & 0.003  & 0.003  & 0.003  & -0.012  & -0.007  \\
          & SE   & 0.034  & 0.017  & 0.022  & 0.040  & 0.049  & 0.038  & 0.047  & 0.046  & 0.028  & 0.032  & 0.035  & 0.094  & 0.077  \\
          & AE   & 0.034  & 0.016  & 0.022  & 0.040  & 0.049  & 0.036  & 0.047  & 0.044  & 0.026  & 0.032  & 0.033  & 0.093  & 0.075  \\
\cmidrule{2-15}          &       &
$\mathbf{A}_{2,33}$ & $\mathbf{B}_{0,21}$ & $\mathbf{B}_{0,22}$ & $\mathbf{B}_{0,31}$ & $\mathbf{B}_{0,32}$ & $\mathbf{B}_{0,33}$ & $\mathbf{B}_{1,11}$ & $\mathbf{B}_{1,22}$ & $\mathbf{B}_{1,33}$ & $\mathbf{B}_{2,11}$ & $\mathbf{B}_{2,22}$ & $\mathbf{B}_{2,33}$ &  \\
\cmidrule{2-15}          & Bias  & -0.009  & 0.010  & 0.011  & -0.002  & 0.002  & -0.004  & 0.002  & 0.003  & 0.004  & -0.010  & -0.013  & -0.009  &  \\
          & SE   & 0.090  & 0.044  & 0.054  & 0.038  & 0.048  & 0.046  & 0.027  & 0.032  & 0.036  & 0.094  & 0.083  & 0.085  &  \\
          & AE   & 0.089  & 0.043  & 0.053  & 0.037  & 0.046  & 0.046  & 0.026  & 0.031  & 0.033  & 0.092  & 0.083  & 0.083  &  \\
    \bottomrule
    \end{tabular}%
  \label{Table_Normal}%
\end{table}%

\begin{table}[htbp]
  \centering
  \caption{The results of $\widehat{\theta}$ when $\mathbf{Z}_t\sim \text{SMT}_{15}$.}
  \scriptsize
    \begin{tabular}{clccccccccccccc}
    \toprule
   $T$ &       & $w$     & $\alpha$ & $\beta$  & $\mathbf{A}_{0,21}$ & $\mathbf{A}_{0,22}$ & $\mathbf{A}_{0,31}$ & $\mathbf{A}_{0,32}$ & $\mathbf{A}_{0,33}$ & $\mathbf{A}_{1,11}$ & $\mathbf{A}_{1,22}$ & $\mathbf{A}_{1,33}$ & $\mathbf{A}_{2,11}$ & $\mathbf{A}_{2,22}$\\
    \midrule
    1000 & Bias  & 0.024  & -0.003  & -0.008  & 0.025  & 0.031  & -0.007  & 0.005  & -0.016  & 0.007  & 0.010  & 0.011  & -0.021  & -0.024  \\
          & SE   & 0.059  & 0.029  & 0.038  & 0.079  & 0.102  & 0.057  & 0.077  & 0.084  & 0.044  & 0.055  & 0.064  & 0.157  & 0.138  \\
          & AE   & 0.057  & 0.028  & 0.037  & 0.072  & 0.095  & 0.056  & 0.072  & 0.079  & 0.040  & 0.051  & 0.059  & 0.156  & 0.135  \\
\cmidrule{2-15}          &       & $\mathbf{A}_{2,33}$ & $\mathbf{B}_{0,21}$ & $\mathbf{B}_{0,22}$ & $\mathbf{B}_{0,31}$ & $\mathbf{B}_{0,32}$ & $\mathbf{B}_{0,33}$ & $\mathbf{B}_{1,11}$ & $\mathbf{B}_{1,22}$ & $\mathbf{B}_{1,33}$ & $\mathbf{B}_{2,11}$ & $\mathbf{B}_{2,22}$ & $\mathbf{B}_{2,33}$ &  \\
\cmidrule{2-15}          & Bias  & -0.023  & 0.021  & 0.025  & -0.010  & 0.004  & -0.022  & 0.010  & 0.009  & 0.009  & -0.017  & -0.013  & -0.012  &  \\
          & SE   & 0.153  & 0.074  & 0.092  & 0.059  & 0.077  & 0.082  & 0.045  & 0.051  & 0.057  & 0.153  & 0.139  & 0.149  &  \\
          & AE   & 0.153  & 0.075  & 0.090  & 0.056  & 0.072  & 0.081  & 0.041  & 0.049  & 0.053  & 0.148  & 0.137  & 0.145  &  \\
    \midrule
         $T$   &       &
          $w$     & $\alpha$ & $\beta$  & $\mathbf{A}_{0,21}$ & $\mathbf{A}_{0,22}$ & $\mathbf{A}_{0,31}$ & $\mathbf{A}_{0,32}$ & $\mathbf{A}_{0,33}$ & $\mathbf{A}_{1,11}$ & $\mathbf{A}_{1,22}$ & $\mathbf{A}_{1,33}$ & $\mathbf{A}_{2,11}$ & $\mathbf{A}_{2,22}$\\
    \midrule
    2000 & Bias  & 0.009  & -0.002  & -0.002  & 0.014  & 0.017  & -0.001  & 0.007  & -0.002  & 0.004  & 0.005  & 0.005  & -0.002  & -0.010  \\
          & SE   & 0.039  & 0.020  & 0.026  & 0.050  & 0.059  & 0.041  & 0.049  & 0.051  & 0.031  & 0.038  & 0.040  & 0.100  & 0.092  \\
          & AE   & 0.040  & 0.020  & 0.026  & 0.050  & 0.058  & 0.039  & 0.048  & 0.050  & 0.030  & 0.037  & 0.039  & 0.097  & 0.089  \\
\cmidrule{2-15}         &       & $\mathbf{A}_{2,33}$ & $\mathbf{B}_{0,21}$ & $\mathbf{B}_{0,22}$ & $\mathbf{B}_{0,31}$ & $\mathbf{B}_{0,32}$ & $\mathbf{B}_{0,33}$ & $\mathbf{B}_{1,11}$ & $\mathbf{B}_{1,22}$ & $\mathbf{B}_{1,33}$ & $\mathbf{B}_{2,11}$ & $\mathbf{B}_{2,22}$ & $\mathbf{B}_{2,33}$ &  \\
\cmidrule{2-15}          & Bias  & -0.015  & 0.012  & 0.014  & -0.002  & 0.003  & -0.006  & 0.005  & 0.007  & 0.007  & -0.008  & -0.011  & -0.009  &  \\
          & SE   & 0.107  & 0.048  & 0.058  & 0.040  & 0.050  & 0.052  & 0.032  & 0.038  & 0.041  & 0.103  & 0.089  & 0.097  &  \\
          & AE   & 0.108  & 0.049  & 0.056  & 0.038  & 0.048  & 0.051  & 0.032  & 0.037  & 0.039  & 0.100  & 0.088  & 0.098  &  \\
    \bottomrule
    \end{tabular}%
  \label{Table_T15}%
\end{table}%

\begin{table}[htbp]
  \centering
  \caption{The results of $\widehat{\theta}$ when $\mathbf{Z}_t\sim \text{SMT}_{25}$.}
  \scriptsize
    \begin{tabular}{clccccccccccccc}
    \toprule
    $T$&       & $w$     & $\alpha$ & $\beta$  & $\mathbf{A}_{0,21}$ & $\mathbf{A}_{0,22}$ & $\mathbf{A}_{0,31}$ & $\mathbf{A}_{0,32}$ & $\mathbf{A}_{0,33}$ & $\mathbf{A}_{1,11}$ & $\mathbf{A}_{1,22}$ & $\mathbf{A}_{1,33}$ & $\mathbf{A}_{2,11}$ & $\mathbf{A}_{2,22}$\\
    \midrule
    1000 & Bias  & 0.022  & -0.009  & -0.006  & 0.023  & 0.026  & -0.009  & 0.006  & -0.016  & 0.008  & 0.009  & 0.009  & -0.013  & -0.017  \\
          & SE   & 0.057  & 0.026  & 0.035  & 0.076  & 0.087  & 0.058  & 0.072  & 0.075  & 0.042  & 0.051  & 0.057  & 0.144  & 0.137  \\
          & AE   & 0.054  & 0.026  & 0.034  & 0.071  & 0.081  & 0.056  & 0.070  & 0.074  & 0.039  & 0.049  & 0.052  & 0.141  & 0.133  \\
\cmidrule{2-15}          &       & $\mathbf{A}_{2,33}$ & $\mathbf{B}_{0,21}$ & $\mathbf{B}_{0,22}$ & $\mathbf{B}_{0,31}$ & $\mathbf{B}_{0,32}$ & $\mathbf{B}_{0,33}$ & $\mathbf{B}_{1,11}$ & $\mathbf{B}_{1,22}$ & $\mathbf{B}_{1,33}$ & $\mathbf{B}_{2,11}$ & $\mathbf{B}_{2,22}$ & $\mathbf{B}_{2,33}$ &  \\
\cmidrule{2-15}          & Bias  & -0.016  & 0.022  & 0.029  & -0.011  & 0.002  & -0.017  & 0.005  & 0.006  & 0.006  & -0.021  & -0.020  & -0.022  &  \\
          & SE   & 0.148  & 0.075  & 0.096  & 0.062  & 0.079  & 0.080  & 0.042  & 0.051  & 0.055  & 0.152  & 0.140  & 0.150  &  \\
          & AE   & 0.142  & 0.071  & 0.093  & 0.057  & 0.072  & 0.080  & 0.038  & 0.047  & 0.051  & 0.147  & 0.136  & 0.148  &  \\
    \midrule
         $T$ &       &
          $w$     & $\alpha$ & $\beta$  & $\mathbf{A}_{0,21}$ & $\mathbf{A}_{0,22}$ & $\mathbf{A}_{0,31}$ & $\mathbf{A}_{0,32}$ & $\mathbf{A}_{0,33}$ & $\mathbf{A}_{1,11}$ & $\mathbf{A}_{1,22}$ & $\mathbf{A}_{1,33}$ & $\mathbf{A}_{2,11}$ & $\mathbf{A}_{2,22}$\\
    \midrule
    2000 & Bias  & 0.010  & -0.002  & -0.003  & 0.012  & 0.015  & -0.002  & 0.003  & -0.004  & 0.004  & 0.004  & 0.005  & -0.008  & -0.009  \\
          & SE   & 0.037  & 0.019  & 0.024  & 0.046  & 0.058  & 0.040  & 0.052  & 0.052  & 0.029  & 0.036  & 0.040  & 0.101  & 0.083  \\
          & AE   & 0.037  & 0.018  & 0.024  & 0.047  & 0.056  & 0.039  & 0.052  & 0.051  & 0.028  & 0.034  & 0.039  & 0.099  & 0.081  \\
\cmidrule{2-15}          &       & $\mathbf{A}_{2,33}$ & $\mathbf{B}_{0,21}$ & $\mathbf{B}_{0,22}$ & $\mathbf{B}_{0,31}$ & $\mathbf{B}_{0,32}$ & $\mathbf{B}_{0,33}$ & $\mathbf{B}_{1,11}$ & $\mathbf{B}_{1,22}$ & $\mathbf{B}_{1,33}$ & $\mathbf{B}_{2,11}$ & $\mathbf{B}_{2,22}$ & $\mathbf{B}_{2,33}$ &  \\
\cmidrule{2-15}          & Bias  & -0.010  & 0.012  & 0.005  & -0.001  & 0.004  & -0.004  & 0.004  & 0.005  & 0.005  & -0.007  & -0.011  & -0.010  &  \\
          & SE   & 0.093  & 0.045  & 0.057  & 0.040  & 0.049  & 0.051  & 0.030  & 0.035  & 0.039  & 0.101  & 0.086  & 0.097  &  \\
          & AE   & 0.091  & 0.044  & 0.056  & 0.038  & 0.048  & 0.049  & 0.028  & 0.033  & 0.039  & 0.098  & 0.084  & 0.097  &  \\
    \bottomrule
    \end{tabular}%
  \label{Table_T25}%
\end{table}%

\subsection{Simulation for the testing}
In this subsection, we exam the finite-sample performance of the portmanteau test $Q_T(L)$. First, we generate 1000 replications of sample size $T=1000$, $2000$, and $4000$ from the following matrix GARCH model
\begin{equation*}
    \mathbf{X}_t = \mathbf{U}_t^{1/2} \mathbf{Z}_t \mathbf{V}_t^{1/2} ,
\end{equation*}
where
\begin{align*}
    \mathbf{U}_t &= \frac{ \mathbf{S}_{1t} }{ \text{tr}(\mathbf{S}_{1t} )  } y_t,\,\,\,\,\mathbf{V}_t = \frac{ \mathbf{S}_{2t} }{ \text{tr}(\mathbf{S}_{2t} )  },\\
  \mathbf{S}_{1t} &= \mathbf{A}_0 \mathbf{A}_0' + \mathbf{A}_1\mathbf{X}_{t-1} \mathbf{X}_{t-1}' \mathbf{A}_1' + \mathbf{A}_2 \mathbf{S}_{1,t-1} \mathbf{A}_2' + \mathbf{A}_3\mathbf{X}_{t-2} \mathbf{X}_{t-2}' \mathbf{A}_3' + \mathbf{A}_4 \mathbf{S}_{1,t-2} \mathbf{A}_4', \\
\mathbf{S}_{2t} &= \mathbf{B}_0 \mathbf{B}_0' + \mathbf{B}_1\mathbf{X}_{t-1}^{'} \mathbf{X}_{t-1} \mathbf{B}_1' + \mathbf{B}_2 \mathbf{S}_{2,t-1} \mathbf{B}_2' + \mathbf{B}_3\mathbf{X}_{t-2}^{'} \mathbf{X}_{t-2} \mathbf{B}_3' + \mathbf{B}_4 \mathbf{S}_{2,t-2} \mathbf{B}_4', \\
y_t &= w + \alpha_1 \text{tr}(\mathbf{X}_{t-1} \mathbf{X}_{t-1}') + \beta_1 y_{t-1} + \alpha_2 \text{tr}(\mathbf{X}_{t-2} \mathbf{X}_{t-2}') + \beta_2 y_{t-2}
\end{align*}
with the parameters $\mathbf{A}_0$, $\mathbf{A}_1$, $\mathbf{B}_0$, and $\mathbf{B}_1$ chosen as before and the remaining parameters
\begin{align*}
w &= 0.4, \alpha_1 = 0.3, \beta_1 = 0.3,\alpha_2 = \delta_1, \beta_2 = \delta_2, \\
\mathbf{A}_2 &= \mathbf{B}_2=\begin{pmatrix} 0.3 & 0 & 0 \\ 0 & 0.3 &0 \\
0 & 0& 0.3\end{pmatrix},\quad
\mathbf{A}_3 = \mathbf{B}_3  = \begin{pmatrix} \delta_1 & 0 & 0 \\ 0 & \delta_1 &0 \\
0 & 0& \delta_1\ \end{pmatrix},\quad
\mathbf{A}_4 = \mathbf{B}_4 = \begin{pmatrix} \delta_2 & 0 & 0 \\ 0 & \delta_2 &0 \\
0 & 0& \delta_2\ \end{pmatrix},
\end{align*}%
and  $\{\mathbf{Z}_t\}_{t=1}^{T}$ is an i.i.d. sequence with $\mathbf{Z}_t\sim\text{MN}(\mathbf{0}, \mathbf{I}_m, \mathbf{I}_n)$. Here, we choose
the values of $\delta_1$ and $\delta_2$ as follows:
\par Case 1: $\delta_1 = 0.038d$ and $\delta_2 = 0$;
\par Case 2: $\delta_1 = 0$ and $\delta_2 = 0.038d$,\\
where $d = 0, 1, ..., 10$. For each case, we take the null (or alternative) model with respect to $d=0$ (or $d\not=0$).

\begin{figure}[!h]
	\centering
	\includegraphics[height=9cm, width=17cm]{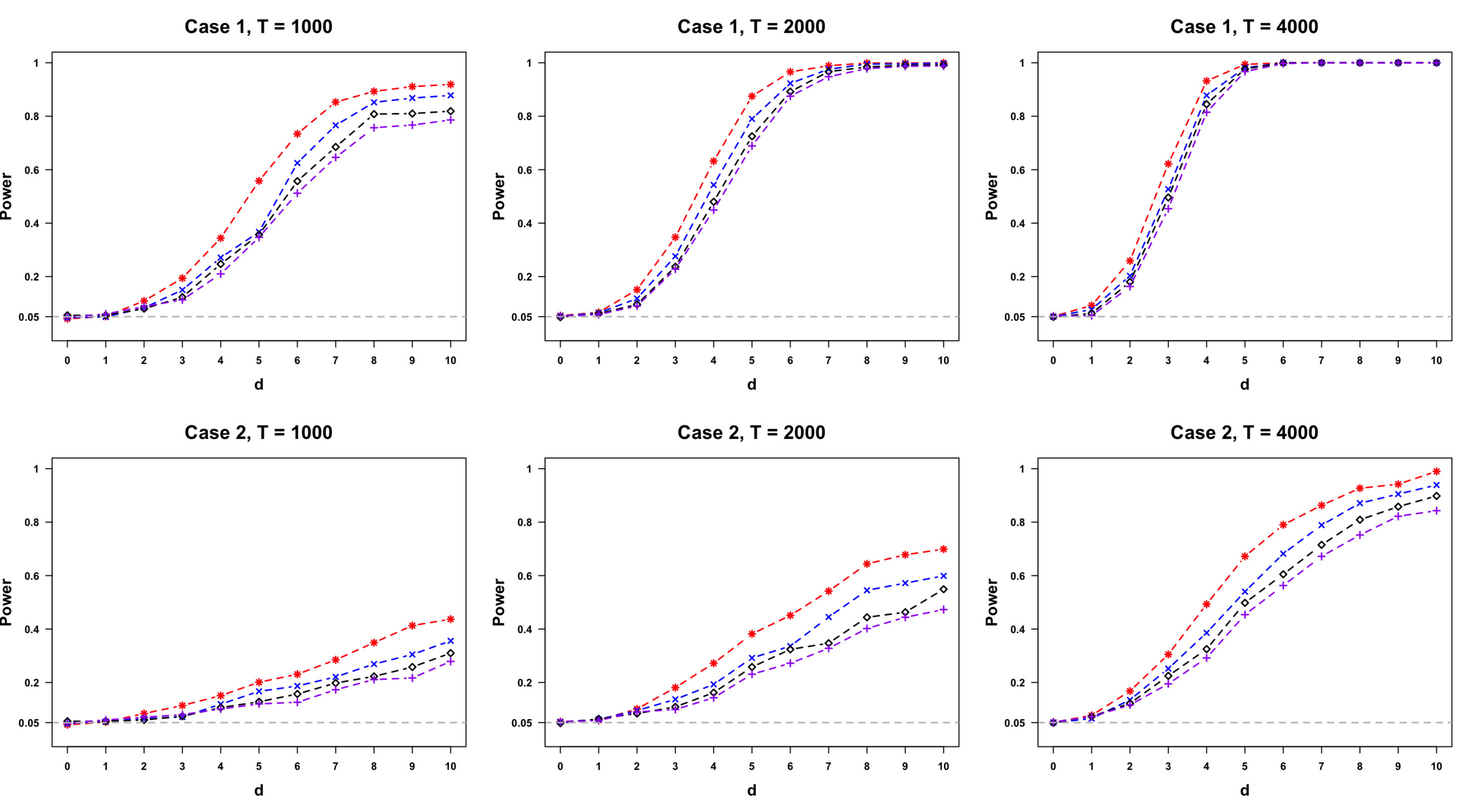}
	\caption{Power across $d$ for $Q_T(L)$ with $L=2$ (star `$\star$' marker), $L=4$ (cross `$\times$' marker), $L=6$ (diamond `$\diamond$' marker), and $L=8$ (plus `$+$' marker). The horizontal dash-dotted line corresponds to the level $5\%$. Upper Panels: Case 1; Bottom Panels: Case 2.}
	\label{Power_plot}
\end{figure}

Next, we fit each replication by the null model, and then apply the portmanteau test $Q_T(L)$ to detect whether the fitted model is adequate. Based on 1000 replications, the empirical power of $Q_T(L)$ across the value of $d$ at the level $\alpha^* = 5\%$ is plotted in Figure \ref{Power_plot}, where the lag $L = 2, 4, 6$, and $8$. From Figure \ref{Power_plot}, we find that (i) all portmanteau tests have precise sizes; (ii) the power of all portmanteau tests becomes large as the value of $T$ or $d$ increases; (iii) $Q_T(2)$ is generally more powerful than other three portmanteau tests; (iv) all portmanteau tests are more powerful to detect the mis-specification of ARCH part in Case 1 than that of GARCH part in Case 2.
In summary, the portmanteau test $Q_T(L)$ performs well especially when $T$ is large.

\section{Real Data Analysis}
\label{sec.real}
In this section, we demonstrate the usefulness  of the matrix GARCH model and matrix factor GARCH model by three applications.
Application 1 applies the matrix GARCH model to study the conditional row and column covariance matrices of a $3\times 3$ matrix time series of Credit Default Swap (CDS) returns.
Applications 2 and 3 adopt the matrix factor GARCH model to investigate  the portfolio allocations in global stock market and China future market, respectively.

\subsection{Application 1}
\begin{figure}[htbp]
	\centering
	\includegraphics[height=9cm, width=15cm]{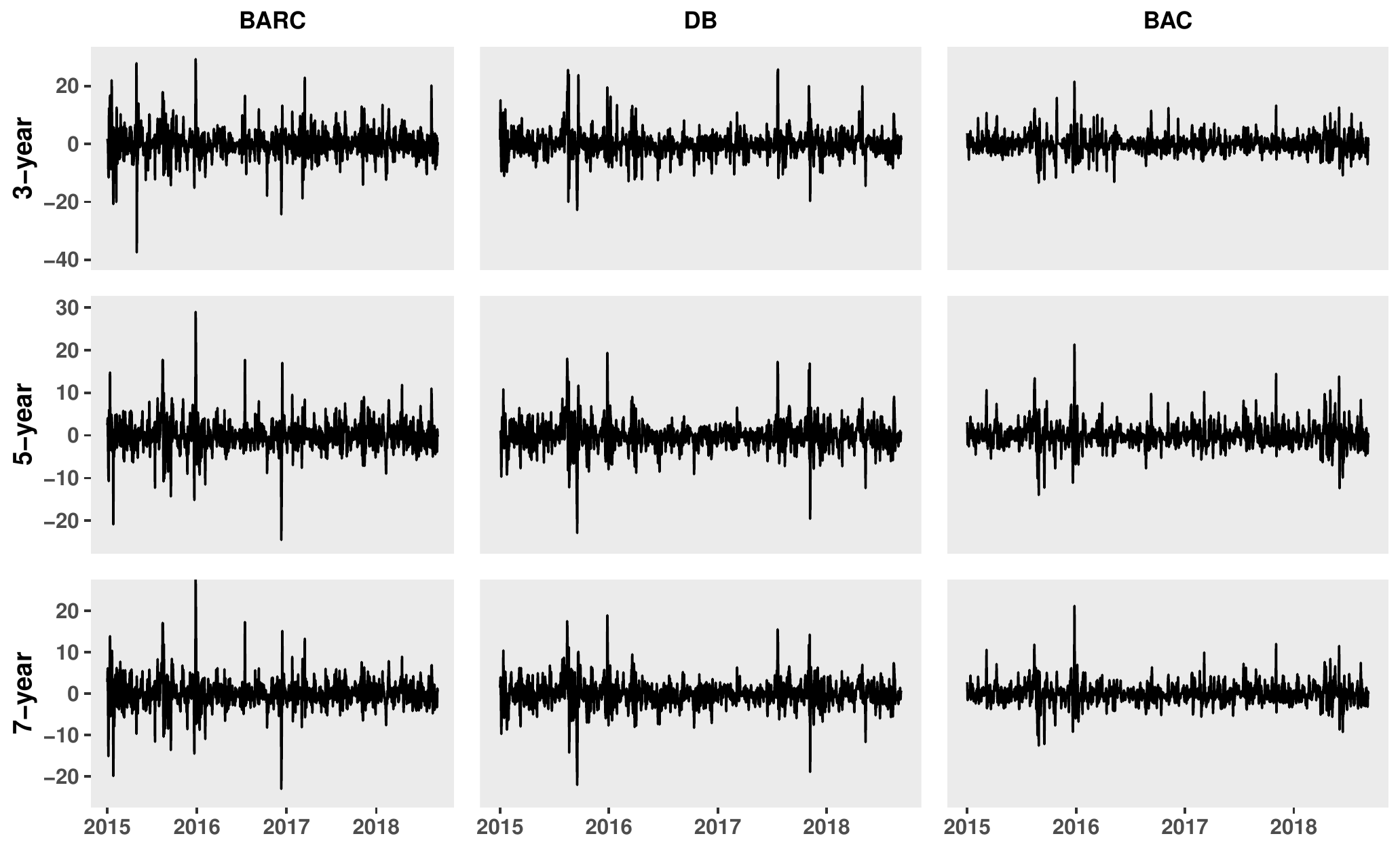}
	\caption{Plot of $3\times 3$ return matrix sequence $\{\mathbf{X}_t\}_{t=1}^{1017}$.}
	\label{Return_Plot}
\end{figure}

In this application, we consider a $3\times 3$ matrix time series, consisting of the daily log-returns (in percentage) of CDS for three financial institutions: Deutsche Bank (DB), Bank of America (BAC), Barclays Bank (BARC) (with respect to columns) in
three tenors of 3 years, 5 years, and 7 years (with respect to rows). The return data for each company and tenor range from $2015/01/01$ to $2018/10/31$, having  $T = 1017$ observations in total. As the sample autocorrelations of each entry series are insignificant, we directly fit the demeaned sequence
 $\{\mathbf{X}_t\}_{t=1}^{1017}$ of this matrix time series (see Figure \ref{Return_Plot} for its visualization) using the matrix GARCH model.

\begin{figure}[htbp]
	\centering
	\includegraphics[height=9cm, width=15cm]{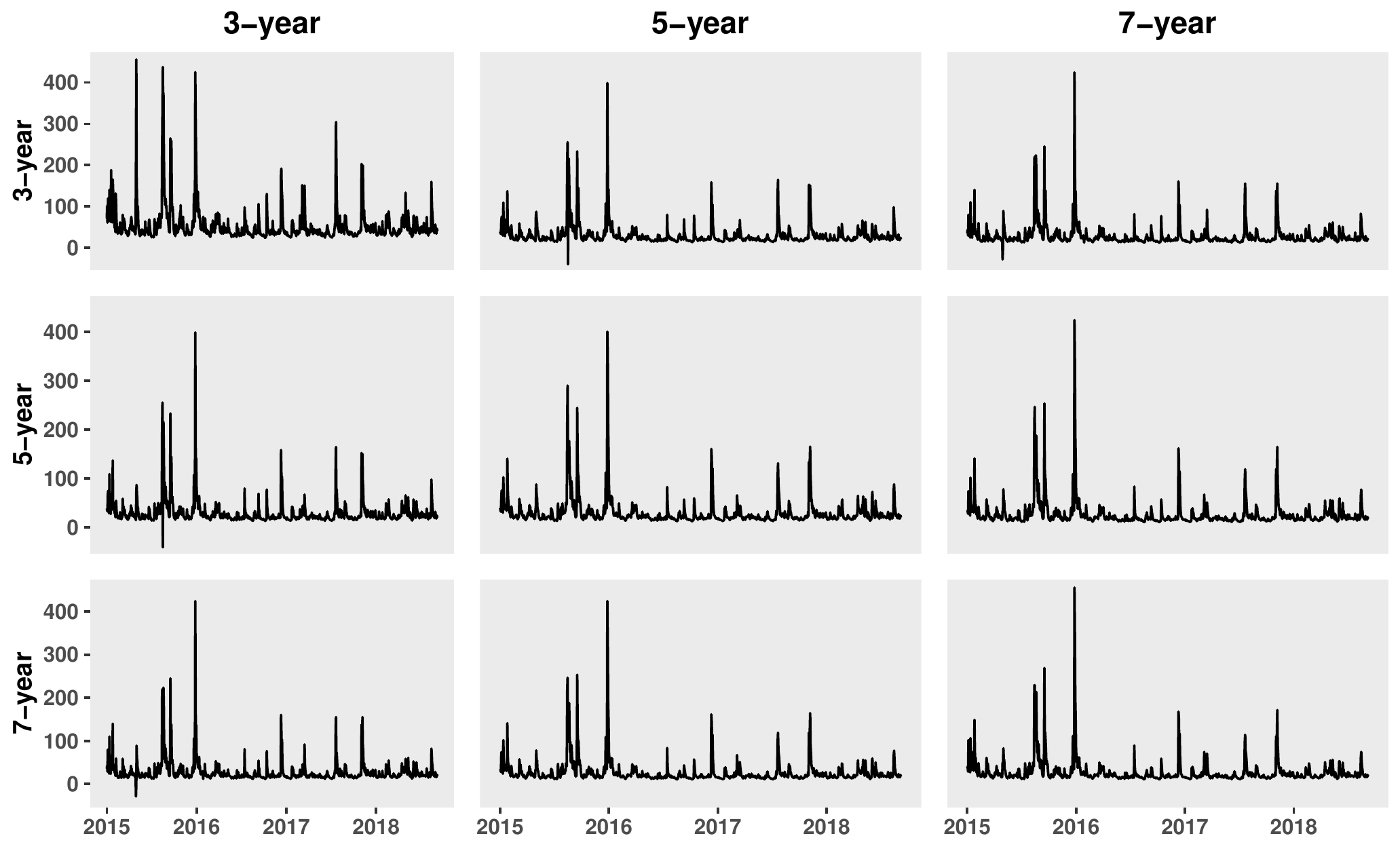}
	\caption{Estimated conditional row covariance matrices $\{\widehat{\mathbf{U}}_t\}_{t=1}^{1017}$.}
	\label{U_Plot}
\end{figure}

\begin{figure}[htbp]
	\centering
	\includegraphics[height=9cm, width=15cm]{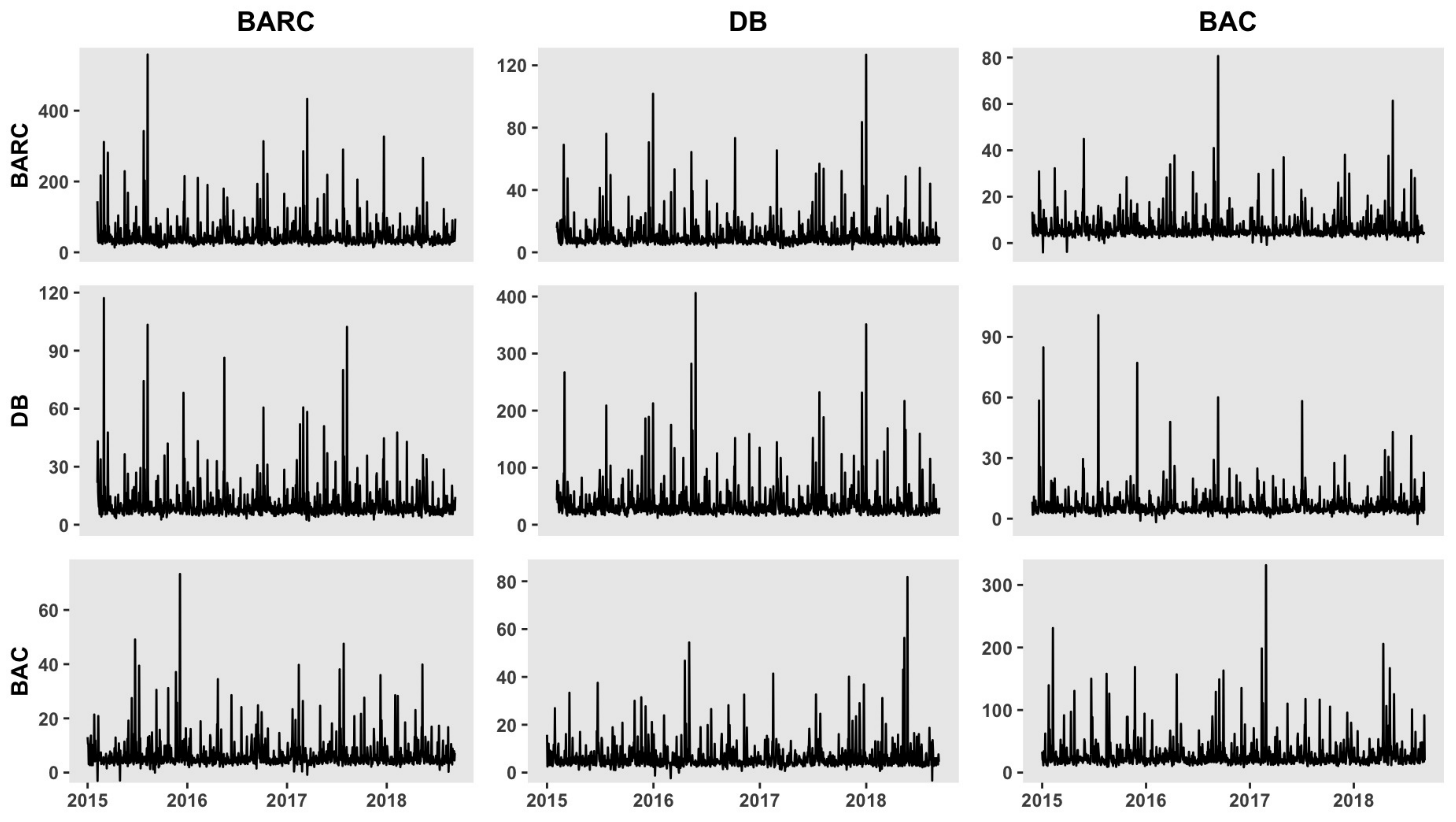}
	\caption{Estimated conditional column covariance matrices $\{\widehat{y}_t\widehat{\mathbf{V}}_t\}_{t=1}^{1017}$.}
	\label{V_Plot}
\end{figure}

Table \ref{Table3} reports the QMLE and corresponding standard errors for the fitted matrix GARCH model. The portmanteau tests $Q_T(2)$, $Q_T(4)$, and $Q_T(6)$ have p-values of 0.476, 0.692, and 0.417, respectively, indicating the adequacy of the fitted model.  Based on the fitted matrix GARCH model, we plot the estimated conditional row and column covariance matrices in Figures \ref{U_Plot} and \ref{V_Plot}, respectively. As expected, we observe from these figures that  (i) each entry of both conditional row and column covariance matrices exhibits the clustering phenomenon over time, and  (ii) in most of times, each entry of conditional row and column covariance matrices has positive values, indicating that the returns of CDS are generally moving together. Meanwhile, we also make some unexpected but interesting observations from these figures.  First, the conditional row covariance matrices generally
are much more stable than the conditional column covariance matrices. Second, the conditional row variance of each tenor has several spikes simultaneously around years 2015 and 2016, and similar spikes also largely exist in the conditional row covariances between different tenors. However, this informative phenomenon cannot be found from the conditional column covariance matrices.
These two findings indicate that the returns of CDS in the same tenor commonly do not usually exhibit large  change for all banks, unless there is a systematic risk in the bank system that causes the CDS price in all tenors to uplift during a very short time period. This kind of systematic risk clearly cannot be captured  solely by examining at the conditional covariance of the CDS returns in different tenors across banks.

\begin{table}[!h]
  \centering
  \caption{The values of QMLE and its standard error.}
  \setlength{\tabcolsep}{5.5mm}{
    \begin{tabular}{ccccc}
    \toprule
    Parameter &       & \multicolumn{3}{c}{$\text{QMLE}_{\text{(standard error)}}$ } \\
    \midrule
    ($\omega$, $\alpha$, $\beta$) &       & $18.9998_{(1.6144)}$ & $0.2385_{(0.018)}$ & $0.5713_{(0.0251)}$ \\
    \midrule
    \multirow{3}[2]{*}{$\mathbf{A}_0$} &       & 1     &       &  \\
          &       & $0.6061_{(0.0318)}$ & $0.5388_{(0.0254)}$ &  \\
          &       & $0.2544_{(0.0702)}$ & $0.2646_{(0.0427)}$ & $-0.2389_{(0.0148)}$ \\
    \midrule
    ($\mathbf{A}_{1,11}$,$\mathbf{A}_{1,22}$,$\mathbf{A}_{1,33}$)    &       &$0.0622_{(0.0039)}$ & $0.0624_{(0.0034)}$ & $0.0702_{(0.0036)}$ \\
    \midrule
    ($\mathbf{A}_{2,11}$,$\mathbf{A}_{2,22}$,$\mathbf{A}_{2,33}$)    &       &$0.1374_{(0.0647)}$ & $-0.4392_{(0.0482)}$ & $-0.3381_{(0.0473)}$ \\
    \midrule
    \multirow{3}[2]{*}{$\mathbf{B}_0$} &       & 1     &       &  \\
          &       & $0.2066_{(0.0243)}$ & $0.1281_{(0.0147)}$ &  \\
          &       & $-0.5713_{(0.0700)}$ & $-0.1662_{( 0.0346)}$ & $0.7038_{(0.0189)}$ \\
    \midrule
    ($\mathbf{B}_{1,11}$,$\mathbf{B}_{1,22}$,$\mathbf{B}_{1,33}$)   &       &$0.0244_{(0.0047)}$ & $0.0385_{(0.0044)}$ & $0.0609_{(0.0055)}$ \\
    \midrule
    ($\mathbf{B}_{2,11}$,$\mathbf{B}_{2,22}$,$\mathbf{B}_{2,33}$)    &       &$0.0721_{(0.1393)}$ & $0.6929_{(0.0752)}$ & $-0.0421_{(0.0815)}$\\
    \bottomrule
    \end{tabular}%
    }
  \label{Table3}%
\end{table}%

Next, we apply the matrix GARCH model to forecast the volatility of all nine entry series $\mathbf{X}_{t,ij}$, where $i,j=1, 2, 3$.
We choose the first $T_{train}=900$ matrix observations as the training data to estimate model parameters and then fix those estimated parameters to do volatility prediction for the testing data that contain the remaining $T_{test}=100$ matrix observations. In addition, we also compare with  the  forecasts  based on the (first order) univariate GARCH model, BEKK model, column BEKK model, and row BEKK model. Specifically, the univariate GARCH model fits each series $\mathbf{X}_{t,ij}$ independently;
the BEKK model fits the $9$-dimensional vector $\textrm{vec}(\mathbf{X}_t)$; the column BEKK model fits each $3$-dimensional
column vector $(\mathbf{X}_{t,1j},\mathbf{X}_{t,2j},\mathbf{X}_{t,3j})'$ independently; and
the row BEKK model fits each $3$-dimensional
row vector $(\mathbf{X}_{t,i1},\mathbf{X}_{t,i2},\mathbf{X}_{t,i3})'$ independently.
All the competing models are fitted by the QMLE method, except for the BEKK model that is estimated by the variance targeted QMLE method in \cite{Pedersen2014TarBEKK} with the diagonal ARCH and GARCH parameter matrices.

After evaluating the forecasting result on the testing data, we summarize the values of mean squared error (MSE), mean absolute error (MAE), and quasi-likelihood (QLIKE) for all considered five models in Table \ref{Table4}. To determine whether the matrix GARCH model provides significantly more accurate volatility predictions than its competitors, we also perform the DM test (\citealp{diebold2002comparing}).
We find from the results of DM test that (i) under the criterions of MSE and MAE, the matrix GARCH model can give more accurate volatility predictions than each competing model, and (ii) under the criterion of QLIKE,  the matrix GARCH model does not have a significant prediction advantage over the BEKK and column BEKK models, though it can significantly outperform the two remaining competing models.
Overall, the matrix GARCH model has demonstrated better forecasting abilities than
 its competing models. This is probably because the
matrix GARCH model makes use of the complete matrix structure of $\mathbf{X}_t$  while avoiding the inclusion of too many parameters.

\begin{table}[htbp]
	\centering
	\begin{threeparttable}
		\caption{Out of sample forecasting comparison.}
  \setlength{\tabcolsep}{4.5mm}{
		\begin{tabular}{cccccc}
			\toprule
            &Matrix
            &Univariate
            &\multirow{2}[1]{*}{BEKK}
            &Column
            &Row \\
			& GARCH & GARCH &  & BEKK & BEKK \\
			\midrule
			MSE   & $7544.859$ & $7770.649^{**}$ & $7622.046^{**}$ & $7855.446^{**}$ & $7887.738^{***}$ \\
			MAE   & $114.94$ & $123.215^{***}$ & $133.919^{***}$ & $124.683^{***}$ & $126.908^{***}$ \\
			QLIKE & $29.570$ & $30.110^{*}$ & $29.807$ & $29.839$ & $30.308^{**}$ \\
			\bottomrule
		\end{tabular}%
}
		\label{Table4}%
		\begin{tablenotes}
			\footnotesize
			\item Note: The result of each competing model is marked with one star, two stars or three stars, if the DM test implies that the matrix GARCH model gives significantly more accurate volatility predictions than this competing model at the level $10 \%$, $5 \%$, or $1 \%$, respectively.
		\end{tablenotes}
	\end{threeparttable}
\end{table}%

\subsection{Application 2}

In this application, we apply the matrix factor GARCH (MF-GARCH) model to perform global portfolio selections across international stock markets.
Specifically, we consider a $10\times 10$ matrix time series, consisting of ten different daily sector indexes (with respect to rows) in ten different stock markets (with respect to columns) from $2017/01/03$ to $2019/11/18$. Here, the  sector indexes are Financials, Energy, Industrials, Materials, Information Technology, Telecommunication Service, Utilities, Consumer Staples, Consumer Discretionary, and Healthcare, while the stock markets are SHSZ in China, HSCI in Hong Kong, KOSPI in Korea, SPX in America, SPTSX in Canada, AS in Australia, SBF in France, NZSE in New Zealand, OSEAX in Norway, and SDEURO in Europe. See a visualization of this daily matrix time series in Figure \ref{Matrix_Index}.
For this matrix time series, we analyze its daily simple returns (in percentage), and denote this return matrix  time series (after demean) by $\{\mathbf{X}_t\}_{t=1}^{700}$ that exhibits no dynamic structure in conditional mean based on its insignificant sample autocorrelations of each entry series. Moreover, we split the entire time period into two parts: the training period containing the first $T_{train}=600$ timepoints and
the testing period including the remaining $T_{test}=100$ timepoints.
At each timepoint $t_0$ in the testing period, we fit the MF-GARCH model based on the latest $T_{train}$ observations $\{\mathbf{X}_t\}_{t=t_0-T_{train}}^{t_0-1}$, and then use this fitted model to predict the conditional covariance matrix of $\textrm{vec}(\mathbf{X}_{t_0})$ by $\widehat{\mathbf{\Sigma}}_{t_0}$, where $\widehat{\mathbf{\Sigma}}_{t_0}$ is computed as in (\ref{est_sigma_x}). Note that we choose $k_1=k_2=3$ in all fitted MF-GARCH models, and this choice of $k_1$ and $k_2$ is guided by the eigenvalue ratio method  (\citealp{YU2022projected}) based on the training data.

Using $\widehat{\Sigma}_{t_0}$, we select the unconstrained minimum variance portfolio (MVP) from 100 sector indexes by choosing the weight vector as
\begin{align}\label{c_MVP}
&\widehat{w}_{t_0}^{u}:=\min_{w_{t_0}'\mathbf{1}_{mn} = 1} w_{t_0}'\widehat{\mathbf{\Sigma}}_{t_0}w_{t_0}=\frac{\widehat{\mathbf{\Sigma}}_{t_0}^{-1}\mathbf{1}_{mn}}{\mathbf{1}_{mn}' \widehat{\mathbf{\Sigma}}_{t_0}^{-1}\mathbf{1}_{mn}}
\end{align}
for $m=n=10$. If the short sales are not preferred or allowed, we can also select the constrained MVP by choosing the weight vector as
\begin{align}\label{u_MVP}
&\widehat{w}_{t_0}^{c}:=\min_{w_{t_0}'\mathbf{1}_{mn} = 1,\,\,w_{t_0}\geq \mathbf{0}} w_{t_0}'\widehat{\mathbf{\Sigma}}_{t_0}w_{t_0},
\end{align}
where $\widehat{w}_{t_0}^{c}$ has no closed form and need be computed by numerical optimization methods.
To evaluate the performance of the proposed unconstrained and constrained MVPs, we follow \cite{Engle2019Large} to
consider the out-of-sample averaged returns (AV), standard deviation of returns (SD), and information ratio (IR) defined by
\begin{align*}
\mbox{AV} = \frac{1}{T_{test}}\sum_{t_0}\widehat{w}_{t_0}'R_{t_0},\,\,\,
\mbox{SD} = \frac{1}{T_{test}-1}\sum_{t=1}^{T_{test}}(\widehat{w}_{t_0}'R_{t_0} - \mbox{AV})^2,\,\,\, \mbox{ and }\,\,\,
\mbox{IR}  = \frac{\mbox{AV}}{\mbox{SD}},
\end{align*}
respectively, where $\widehat{w}_{t_0}$ is $\widehat{w}_{t_0}^{u}$ (or $\widehat{w}_{t_0}^{c}$) with respect to unconstrained (or constrained) MVP, and $R_{t_0}=\mathrm{vec}(\mathbf{X}_{t_0})$. Certainly, many other GARCH-type models can also be used to predict the conditional covariance matrix of $\textrm{vec}(\mathbf{X}_{t_0})$, and  the  related unconstrained and constrained MVPs can be constructed similarly as done in (\ref{c_MVP})--(\ref{u_MVP}). For  comparison purpose,  we report the results based on  the generalized orthogonal GARCH (GO-GARCH) model (\citealp{van2002go}), DCC model (\citealp{engle2002dynamic}),
Risk Metrics model (\citealp{zumbach2007riskmetrics}), and Equal Weights model. For the DCC model, its QMLE may not perform well when
the dimension of $\textrm{vec}(\mathbf{X}_{t_0})$ is 100. Therefore, we further follow \cite{Engle2019Large} to estimate the DCC model
by using either the linear shrinkage (LS) method in \cite{ledoit2004well} or nonlinear shrinkage (NLS) method in \cite{ledoit2012nonlinear}, leading to the DCC-LS and DCC-NLS models for comparison.


Table \ref{Table_Portfolio} reports the annualized AV, SD, and IR of the unconstrained and constrained MVPs proposed by the MF-GARCH model and its six competitors. From this table, we find that the unconstrained and constrained MVPs selected by the MF-GARCH model perform the best in terms of AV, SD, and IR. In particular, the values of IR for the unconstrained and constrained MVPs from the MF-GARCH model are 28\% and 20\% higher than those from the DCC-NLS model, which has the best performance among all four competing multivariate GARCH-type models.
The advantage of the MF-GARCH model over its four multivariate GARCH-type competitors is most likely due to its matrix structure, which can largely alleviate the trouble of dimensionality and reduce the computational burden in model estimation.
Moreover, we find that the Risk Metrics model performs the worst. This is expected since a simple weighting parameter used by the Risk Metrics model could not adequately capture the dynamics of $\textrm{vec}(\mathbf{X}_{t})$. Another observation is that
the Equal Weights model has an unsatisfactory performance, indicating
 the necessity of weight selection from the dynamics of conditional covariance matrix of $\textrm{vec}(\mathbf{X}_{t})$.

%

\begin{table}[!htp]
  \centering
  \begin{threeparttable}
  \caption{MVP performance based on global sector index matrix data.}
  \setlength{\tabcolsep}{4.8mm}{
    \begin{tabular}{ccccccccc}
    \toprule
          &       & \multicolumn{3}{c}{Unconstrained MVP} &       & \multicolumn{3}{c}{Constrained MVP} \\
\cmidrule{3-5}\cmidrule{7-9}          &       & AV    & SD    & IR    &       & AV    & SD    & IR \\
    \midrule
    MF-GARCH &       & \pmb{12.79} & \pmb{3.71} & \pmb{3.44} &       & \pmb{12.07} & \pmb{3.43} & \pmb{3.51} \\
    GO-GARCH &       & 7.02  & 4.25  & 1.65  &       & 6.44  & 3.74  & 1.72 \\
    DCC   &       & 5.64  & 3.96  & 1.33  &       & 6.14  & 3.51  & 1.75 \\
    DCC-LS &       & 9.76  & 3.79  & 2.58  &       & 9.69  & 3.45  & 2.81 \\
    DCC-NLS &       & 10.09 & 3.76  & 2.68  &       & 10.08 &3.44  & 2.93 \\
    Risk Metrics &       & -6.41 & 7.97  & -0.81 &       & 5.64  & 4.49  & 1.26 \\
    Equal Weights &       & 4.91  & 3.79  & 1.29  &       & 4.91  & 3.79  & 1.29 \\
    \bottomrule
    \end{tabular}%
    }
  \label{Table_Portfolio}%
  		\begin{tablenotes}
			\footnotesize
			\item Note: The best values of AV, SD, and IR are in boldface.
		\end{tablenotes}
\end{threeparttable}
\end{table}%

\begin{figure}[!h]
	\centering
	\includegraphics[height=18cm, width=15cm]{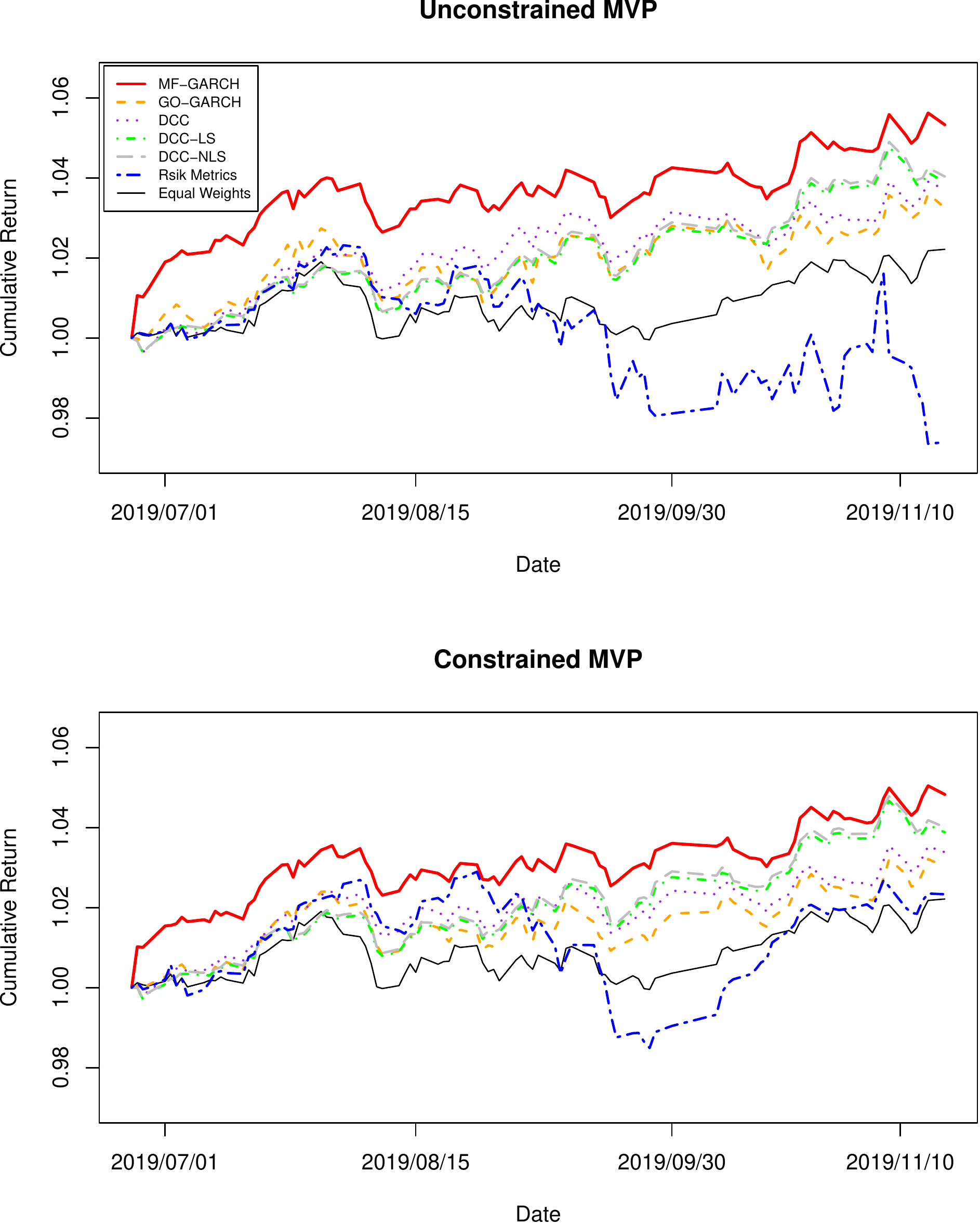}
	\caption{Cumulative returns of MVPs based on global sector index matrix data.}
	\label{Cum_index}
\end{figure}

Last, we plot the out-of-sample cumulative returns of unconstrained and constrained MVPs constructed by all considered models in Figure \ref{Cum_index}. From this figure, we find that  the unconstrained and constrained MVPs selected by the MF-GARCH model have the best cumulative returns over time. This finding is consistent to the one from Table \ref{Table_Portfolio} that the largest value of AV for each MVP is achieved by the MF-GARCH model.

\subsection{Application 3}

\begin{figure}[htbp]
	\centering
	\includegraphics[height=6.5cm, width=17cm]{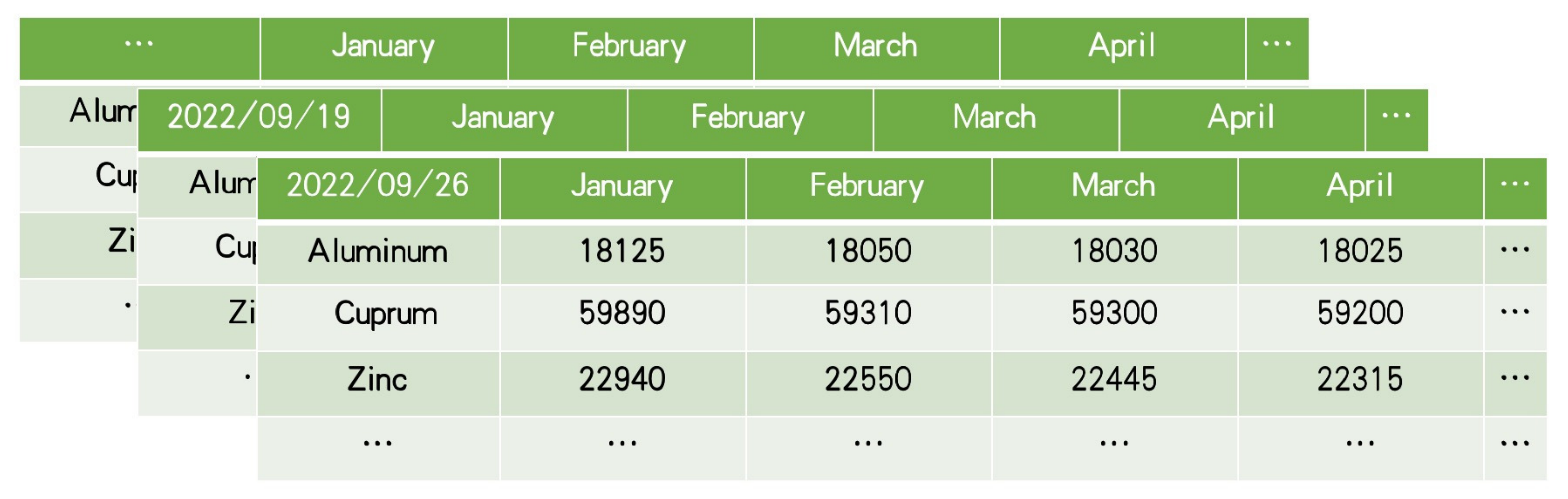}
	\caption{A real example of matrix-variate observations consisting of future prices of varieties with different delivery months. Data source: Commodity Exchange in Shanghai, Dalian, and Zhengzhou.}
	\label{Matrix_Future}
\end{figure}

To further demonstrate the usefulness of the MF-GARCH model, we consider another $8\times 12$ matrix time series that is comprised of weekly closing prices of futures in $8$ varieties (with respect to rows) with $12$ different delivery months (with respect to columns) from $2009/08/19$ to $2022/09/26$, where the varieties include Aluminum, Cuprum, Zinc, Polyethylene, Deformed Steel Bar, Palm Oil, Polyvinyl Chloride, and Pure Terephthalic Acid, and the delivery months are from January to December. See a visualization of this weekly matrix time series in Figure \ref{Matrix_Future}.

Below, we study the weekly simple returns (in percentage) of this matrix time series and denote the corresponding return matrix time series (after demean) by $\{\mathbf{X}_t\}_{t=1}^{637}$. Similar to the implementations in Application 2, we apply the MF-GARCH model and other six competing models to propose the unconstrained and constrained MVPs during the testing period, which contains the last $T_{test}=100$ timepoints.
For the MF-GARCH model, we pick up $k_1=2$ and $k_2=3$ according to the eigenvalue ratio method  (\citealp{YU2022projected}) based on the first $T_{train}=537$ observations.

\begin{figure}[!h]
	\centering
	\includegraphics[height=18cm, width=15cm]{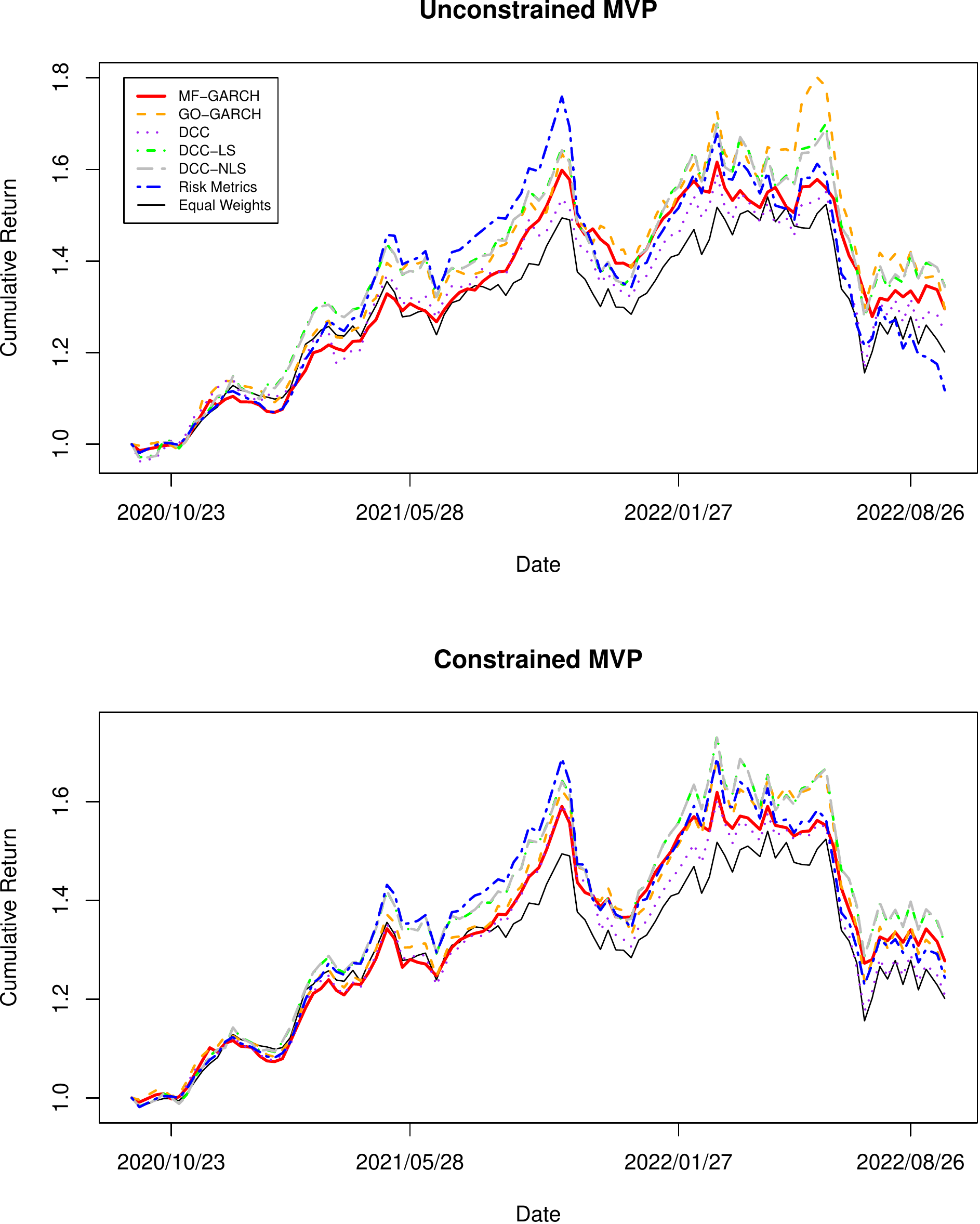}
	\caption{Cumulative returns of MVPs based on future price matrix data.}
	\label{Cum_Future}
\end{figure}

Table \ref{Table_Future} reports the annualized AV, SD, and IR of the unconstrained and constrained MVPs proposed by all considered models.
From this table, we find that the unconstrained and constrained MVPs selected by the MF-GARCH model have the best values of SD and IR, while
those selected by the DCC-LS or DCC-NLS model have the best value of AV. This finding implies that the higher return of MVPs
from the DCC-LS and DCC-NLS models is obtained at the price of higher risk, and the MF-GARCH model having the largest value of IR tends to make a better balance between return and risk. This advantage of MF-GARCH model can also be observed from Figure \ref{Cum_Future}, in which the out-of-sample cumulative returns of two MVPs from the MF-GARCH model seem to be more stable over time than those from other models.


%

\begin{table}[!htbp]
  \centering
    \begin{threeparttable}
  \caption{MVP performance based on future price matrix data.}
   \setlength{\tabcolsep}{4.8mm}{
    \begin{tabular}{ccccccccc}
    \toprule
          &       & \multicolumn{3}{c}{Unconstrained MVP} &       & \multicolumn{3}{c}{Constrained MVP} \\
\cmidrule{3-5}\cmidrule{7-9}          &       & AV    & SD    & IR    &       & AV    & SD    & IR \\
    \midrule
    MF-GARCH &       & 70.33 & \pmb{35.22} & \pmb{1.99} &       & 69.65 & \pmb{39.49} & \pmb{1.77} \\
    GO-GARCH &       & 77.22 & 46.82 & 1.65  &       & 67.13 & 47.88 & 1.41 \\
    DCC   &       & 70.15 & 47.74 & 1.47  &       & 58.61 & 45.71 & 1.28 \\
    DCC-LS &       & \pmb{89.32} & 49.29 & 1.81  &       & 80.74 & 47.98 & 1.68 \\
    DCC-NLS &       & 88.09 & 49.17 & 1.79  &       & \pmb{81.31} & 48.31 & 1.68 \\
    Risk Metrics &       & 42.24 & 51.35 & 0.82  &       & 66.85 & 47.61 & 1.41 \\
    Equal Weights &       & 56.86 & 45.17 & 1.26  &       & 56.85 & 45.17 & 1.26 \\
    \bottomrule
    \end{tabular}%
    }
  \label{Table_Future}%
   		\begin{tablenotes}
			\footnotesize
			\item Note: The best values of AV, SD, and IR are in boldface.
		\end{tablenotes}
    \end{threeparttable}
\end{table}%

%

\section{Concluding Remarks}\label{sec_concluding}

In this paper, we propose a novel matrix GARCH model to study the conditional heteroskedasticity of matrix time series.
This matrix GARCH model is comprised of three specifications, including two BEKK specifications (normalized by their own traces)
to describe the conditional row and column covariance matrices, and one univariate GARCH specification to
capture the trace of conditional row or column covariance matrix. We show that the univariate GARCH specification for the trace is necessary to identify the conditional row and column covariance matrices.
Next, we propose the QMLE for the matrix GARCH model and construct the portmanteau test for the model diagnostic checking.
Under certain conditions, we establish the asymptotics of the QMLE and portmanteau test.
Moreover, we introduce a matrix factor GARCH model to deal with large dimensional matrix time series.
Compared with existing multivariate GARCH-type models that can also study the conditional heteroskedasticity of matrix time series after vectorization, the matrix GARCH and matrix factor GARCH models not only have a clear interpretation on conditional row and column covariance matrices, but also significantly reduce the number of model parameters. These two advantages are extensively demonstrated by three real examples.

 Our matrix GARCH model opens a new door for studying the conditional heteroskedasticity of complex structured time series.
 In line with the vast body of literature on multivariate GARCH-type models, one could consider the extension of our matrix-GARCH model to better capture the heavy-tailedness and asymmetry of matrix time series data. Another interesting
future work is to investigate the conditional heteroskedasticity of tensor-variant time series (\citealp{wang2021high,chen2022factor}) by using a similar approach as our matrix GARCH model. These future works along with the current work, have the potential to generate valuable toolkits for practitioners to analyze complex time series.

\linespread{1.3}\selectfont
\bibliographystyle{elsarticle-harv}
\bibliography{ref}

\end{document}